\documentclass[prd,preprint,floatfix,nofootinbib,superscriptaddress,noshowpacs,showkeys]{revtex4} 
\pdfoutput=1
\usepackage[caption=false]{subfig}
\usepackage{amsmath}
\usepackage{amsfonts}
\usepackage{amssymb}
\usepackage{float}
\usepackage{color}
\usepackage{graphicx}
\usepackage{graphics}
\usepackage{hyperref}
\usepackage{adjustbox,array}
\usepackage{enumitem} 
\hypersetup{}
\usepackage[utf8x]{inputenc} 
\usepackage{epstopdf}
\usepackage{multirow}
\usepackage{slashed}
\usepackage{booktabs}
\usepackage{cancel}
\usepackage{mathrsfs}
\usepackage{diagbox}
\usepackage{csquotes}

\definecolor{rosso}{cmyk}{0,1,1,0.4}
\definecolor{rossos}{cmyk}{0,1,1,0.55}
\definecolor{rossoc}{cmyk}{0,1,1,0.2}
\definecolor{blu}{cmyk}{1,1,0,0.3}
\definecolor{blus}{cmyk}{1,1,0,0.6}
\definecolor{bluc}{cmyk}{1,1,0,0.1}
\definecolor{verde}{cmyk}{0.92,0,0.59,0.25}
\definecolor{verdec}{cmyk}{0.92,0,0.59,0.15}
\definecolor{verdes}{cmyk}{0.92,0,0.59,0.4}

\AtBeginDocument{
\heavyrulewidth=.08em
\lightrulewidth=.05em
\cmidrulewidth=.03em
\belowrulesep=.65ex
\belowbottomsep=0pt
\aboverulesep=.4ex
\abovetopsep=0pt
\cmidrulesep=\doublerulesep
\cmidrulekern=.5em
\defaultaddspace=.5em
}

\hypersetup{
 colorlinks=true,
 linkcolor=verdec,
 urlcolor=verdec,
 citecolor=verdec
}

\newcommand{ \eq}[1]{Eq.~(\ref{#1})}

\newcommand{\gsim}{\gtrsim}
\newcommand{\lsim}{\lesssim}

\newcommand{\lf}{\left(}
\newcommand{\ri}{\right)}

\newcommand{\nn}{\nonumber}

\newcommand{\sqt}{\sqrt{2}}

\renewcommand{\lg}{\mathscr{L}} 

\newcommand{\mco}{\mathcal{O}}

\newcommand{\br}{\mathcal{B}}
\newcommand{\hc}{{\rm H.c.}}

\newcommand{\sm}{{\rm SM}}

\newcommand{\pb}{{\;{\rm pb}}}
\newcommand{\fb}{{\;{\rm fb}}}

\newcommand{\gev}{{\;{\rm GeV}}}
\newcommand{\tev}{{\;{\rm TeV}}}

\newcommand{\beq}{\begin{equation}}
\newcommand{\eeq}{\end{equation}}
\newcommand{\bea}{\begin{eqnarray}}
\newcommand{\eea}{\end{eqnarray}}
\newcommand{\barr}{\begin{array}}
\newcommand{\earr}{\end{array}}
\newcommand{\bc}{\begin{center}}
\newcommand{\ec}{\end{center}}
\newcommand{\bit}{\begin{itemize}}
\newcommand{\eit}{\end{itemize}}
\newcommand{\ben}{\begin{enumerate}}
\newcommand{\een}{\end{enumerate}}

\newcommand{\al}{\alpha}
\newcommand{\bt}{\beta}

\newcommand{\dt}{\delta}
\newcommand{\Dt}{\Delta}

\newcommand{\dma}{\Delta M_A}
\newcommand{\dmhh}{\Delta M_H}
\newcommand{\sg}{\sigma}
\newcommand{\es}{\epsilon}
\newcommand{\kp}{\kappa}

\newcommand{\lmh}{\hat{\lambda}}

\newcommand{\gm}{\gamma}
\newcommand{\Gm}{\Gamma}
\newcommand{\lm}{\lambda}

\newcommand{\Lm}{\Lambda}
\newcommand{\damu}{\Delta a_\mu}

\newcommand{\mwcdf}{m_W^{\rm CDF}}
\newcommand{\hsm}{{h_{\rm SM}}}
\newcommand{\ch}{H^\pm}

\newcommand{\wpm}{W^\pm}

\newcommand{\mh}{m_{h}}
\newcommand{\mhsm}{\mh}
\newcommand{\mch}{M_{H^\pm}}
\newcommand{\mhh}{M_{H}}
\newcommand{\ma}{M_{A}}

\newcommand{\ca}{c_\alpha}
\newcommand{\sa}{s_\alpha}

\newcommand{\tb}{t_\beta}

\newcommand{\cb}{c_\beta}
\renewcommand{\sb}{s_\beta}

\newcommand{\cba}{c_{\beta-\alpha}}
\newcommand{\sba}{s_{\beta-\alpha}}

\newcommand{\lmc}{{\Lambda_{\rm c}}}

\newcommand{\ee}      {{e^+ e^-}}
\newcommand{\mmu}      {{\mu^+ \mu^-}}

\newcommand{\ttau}      {{\tau^+\tau^-}} 

\newcommand{\ttop}      {{t\bar{t}}}

\newcommand{\bb}      {{b \bar{b}}}

\definecolor{mint}{rgb}{0.24, 0.71, 0.54}

\begin{document}

\title{\color{verdes} 
CDF $W$ boson mass and muon $g-2$ \\ in type-X two-Higgs-doublet model
\\with a Higgs-phobic light pseudoscalar }
\author{Jinheung Kim}
\email{jinheung.kim1216@gmail.com}
\address{Department of Physics, Konkuk University, Seoul 05029, Republic of Korea}
\author{Soojin Lee}
\email{soojinlee957@gmail.com}
\address{Department of Physics, Konkuk University, Seoul 05029, Republic of Korea}
\author{Prasenjit Sanyal}
\email{prasenjit.sanyal@apctp.org}
\address{Asia Pacific Center for Theoretical Physics, Pohang 37673, Republic of Korea}
\author{Jeonghyeon Song}
\email{jhsong@konkuk.ac.kr}
\address{Department of Physics, Konkuk University, Seoul 05029, Republic of Korea}

\begin{abstract}
The recent measurement of the $W$ boson mass by the CDF collaboration adds an anomaly to the long-standing discrepancy in the muon anomalous magnetic moment, $\Delta a_\mu$. Although type-X in the two-Higgs-doublet model provides an attractive solution to $\Delta a_\mu$ through a light pseudoscalar $A$, the model confronts the exotic Higgs decays of $h\to AA$ and the lepton flavor universality data in the $\tau$ and $Z$ decays. To save the model, we propose that the light pseudoscalar be Higgs-phobic. Through the random scanning over the entire parameter space, we perform a comparative study of the Higgs-phobic type-X with and without the CDF $m_W$ measurement, called the CDF and PDG cases respectively. Both cases can explain the two anomalies as well as all the other constraints, but have significant differences in the finally allowed parameter space. For example, a small region with almost degenerate masses of new Higgs bosons around 100 GeV is allowed only in the PDG case. The cutoff scale of the model is also studied via the analysis of renormalization group equations, which reaches up to $10^5~{\rm GeV}$ ($10^7~{\rm GeV}$) in the CDF (PDG) case.  Since the dominant decay modes are $A\to \tau\tau$, $H \to Z A$, and $H^\pm\to W^\pm A$ in most of the viable parameter space, we propose the $4\tau+VV'$ states as the golden discovery channel at the LHC.
\end{abstract}

\vspace{1cm}
\keywords{Higgs Physics, Beyond the Standard Model}
\preprint{APCTP Pre2022 - 006}
\maketitle
\tableofcontents

\section{Introduction}
The CDF collaboration at the Fermilab  National Accelerator Laboratory  has come out with the most precise measurement of $W$ boson mass \cite{CDF:2022hxs}
\begin{eqnarray}
m_W^{\rm CDF} = 80.4335 \pm 0.0094 ~ \text{GeV},
\end{eqnarray}
using the data set collected at $8.8\fb^{-1}$ luminosity. The new mass deviates from the Standard Model (SM) prediction of $m^\sm_W = 80.357 \pm 0.006$ GeV~\cite{ParticleDataGroup:2020ssz} by $7\sigma$. 
Previously the world average of $m_{W}$ measurements~\cite{ParticleDataGroup:2020ssz} 
was only $1.8\sigma$ standard deviation from $m^\sm_W$. The discrepancy of the $W$ mass still needs to be confirmed as there is a tension between the CDF measurement and ATLAS report~\cite{ATLAS:2017rzl}. 
However, if we accept the new mass of $W$ boson then the validity of the SM is under serious question.
An efficient way to parameterize the discrepancy of $W$ boson mass is 
the Peskin-Takeuchi oblique parameter ($S$, $T$, and $U$): a new physics model beyond the SM (BSM) can be explored by its contribution to the gauge boson self energies. 
In most models, the contribution to $U$ is significantly small, so setting $U=0$ is usually accepted.
Then we have large shift of the central values~\cite{Lu:2022bgw,Asadi:2022xiy,Balkin:2022glu,Strumia:2022qkt,deBlas:2022hdk}
such that $S_{\rm CDF} = 0.15 \pm 0.08$ and $T_{\rm CDF} = 0.27 \pm 0.06$ with the correlation $\rho_{ST}=0.93$~\cite{Lu:2022bgw}.     
Various BSM models have been studied to explain the new oblique parameters~\cite{Fan:2022dck,Zhu:2022tpr,Lu:2022bgw,Zhu:2022scj,Song:2022xts,Bahl:2022xzi,Heo:2022dey,Babu:2022pdn,Biekotter:2022abc,Ahn:2022xeq,Han:2022juu,Arcadi:2022dmt,Ghorbani:2022vtv,Cheng:2022jyi,Du:2022brr,Kanemura:2022ahw,Mondal:2022xdy,Borah:2022obi,Yang:2022gvz,Du:2022pbp,Athron:2022isz,Zheng:2022irz,Ghoshal:2022vzo,Blennow:2022yfm,Arias-Aragon:2022ats,Liu:2022jdq,Popov:2022ldh,Crivellin:2022fdf,deBlas:2022hdk,Fan:2022yly,Bagnaschi:2022whn,Paul:2022dds,Gu:2022htv,DiLuzio:2022xns,Endo:2022kiw,Balkin:2022glu,Cirigliano:2022qdm,Yuan:2022cpw,Strumia:2022qkt,Cacciapaglia:2022xih,Sakurai:2022hwh,Heckman:2022the,Krasnikov:2022xsi,Peli:2022ybi,Perez:2022uil,Wilson:2022gma,Zhang:2022nnh,Carpenter:2022oyg,Du:2022fqv,Lee:2022gyf,Chen:2022ocr,Cao:2022mif,Abouabid:2022lpg}.

Another long-standing problem in particle physics is the muon anomalous magnetic moment. The combined result of  the Fermilab National Accelerator Laboratory experiment~\cite{Muong-2:2021ojo,Muong-2:2021vma} and the Brookhaven National Laboratory experiment~\cite{Muong-2:2006rrc} has shown 
a deviation from the SM prediction~\cite{Aoyama:2012wk,Aoyama:2019ryr,Czarnecki:2002nt,Gnendiger:2013pva,Davier:2017zfy,Keshavarzi:2018mgv,Colangelo:2018mtw,Hoferichter:2019mqg,Davier:2019can,Keshavarzi:2019abf,Kurz:2014wya,Melnikov:2003xd,Masjuan:2017tvw,Colangelo:2017fiz,Hoferichter:2018kwz,Gerardin:2019vio,Bijnens:2019ghy,Colangelo:2019uex,Blum:2019ugy,Colangelo:2014qya,Aoyama:2020ynm} by $4.2\sigma$, which is reported to be 
\begin{eqnarray}
\label{eq:damu:obs}
\Delta a_\mu = a_{\mu}^{\rm exp} - a_{\mu}^{\rm SM} = 251(59) \times 10^{-11}.
\end{eqnarray} 
Two anomalies of  $m_W^{\rm CDF}$ and $\damu$ call for new physics. 
Several works have been done to simultaneously explain the two anomalies in the context of $U(1)$ gauge extended models with vectorlike leptons~\cite{Lee:2022nqz,Baek:2022agi,Zhou:2022cql}, vector leptoquark model~\cite{Cheung:2022zsb}, scalar leptoquark model~\cite{Bhaskar:2022vgk,Athron:2022qpo}, Zee model~\cite{Chowdhury:2022moc}, vectorlike lepton models~\cite{Kawamura:2022uft,Nagao:2022oin}, a flavor conserving two-Higgs-doublet model (2HDM)~\cite{Botella:2022rte}, and next-to-minimal supersymmetric model~\cite{Tang:2022pxh}.
 
In this paper, we study type-X (lepton-specific) 2HDM in light of
the CDF $W$ boson mass and muon $g-2$ anomalies. 
Type-X has drawn a lot of interest as an explanation of $\damu$~\cite{Abe:2015oca,Han:2015yys,Cherchiglia:2016eui,Cherchiglia:2017uwv,Han:2018znu,Wang:2018hnw,Chun:2019oix,DelleRose:2020oaa,Jana:2020pxx,Ghosh:2020tfq,Jueid:2021avn,Athron:2021evk}.
One of its most salient characteristics is the enhanced coupling of the BSM Higgs bosons (neutral \textit{CP}-even $H$, \textit{CP}-odd $A$, and charged Higgs $H^\pm$) to the leptons by $\tan\beta$, the ratio of the vacuum expectation values of two Higgs doublet fields. 
Through the enhanced leptonic coupling, 
type-X can explain muon $g-2$ anomaly via two loop Barr-Zee diagram with $\tau$-loop~\cite{Barr:1990vd,Ilisie:2015tra}. 
The contributions to $\damu$ 
can be sizable and positive with large $\tan\beta$ and small $\ma$. 
However, a light pseudoscalar with $M_A < m_h^\sm/2$  opens up $h_{SM} \to AA$ which is severely constrained by $h_{SM} \to AA \to 4\tau/2\mu2\tau$ channels \cite{CMS:2018qvj}.
Kinematical solution of $M_A >m_h^\sm/2$
demands very large $\tan\beta$ above $ 100$ for the explanation of $\damu$.
Then this extremely large $\tan\beta$ enhances the contributions to the lepton flavor universality (LFU) data in the $\tau$ and $Z$ decays,
which invalidates the model~\cite{Jueid:2021avn}.
This motivates us to consider the Higgs-phobic type-X where the vertex $h$-$A$-$A$
vanishes.

An essential question is how the changes of $S$ and $T$ due to the CDF $W$ boson mass 
affect the parameter space compatible with the muon $g-2$
as well as all the theoretical and experimental constraints. 
To comprehensively answer the question, we will perform a scan over the entire parameter space
in four steps, considering both the old and new sets of $S$ and $T$. 
In step I, we impose the theoretical bounds (vacuum stability of the potential, unitarity,  perturbativity) 
and the muon $g-2$ constraint. 
In step II, we include the $S$ and $T$ parameters before
and after the CDF $m_W$ measurement. 
In step III, we impose the Higgs precision data and the most updated direct search bounds from the LEP, Tevatron, and LHC. 
In step IV, we further restrict the parameter space 
through the global $\chi^2$ fit to $\damu$ and the LFU data.
Based on the scan results,
we will find the common and different features before and after the CDF $m_W$ measurement.
Another important question 
is to what energy scale the finally allowed parameter points survive.
We will perform the renormalization group equation (RGE) analysis to obtain 
the cutoff scale $\lmc$ of every viable parameter point.
The final question is how to probe the Higgs-phobic type-X at the LHC. 
In the literature, 
the multi-$\tau$ states have extensively been studied for type-X,
$2\tau$~\cite{Cheung:2022ndq},
$2\mu2\tau$~\cite{Kanemura:2011kx,Chun:2017yob},
$\bb\ttau$~\cite{Kanemura:2021dez},
$3\tau$~\cite{Kanemura:2011kx,Chun:2015hsa},
$4\tau$~\cite{Kanemura:2011kx,Chun:2015hsa}, and
$4\tau+V$~\cite{Chun:2015hsa}.
We will show that $4\tau + ZW/WW$ is the golden discovery channel.
These are our new contributions.

  The paper is organized in the following way. 
  In Sec.~\ref{sec:review}, we give a brief review of type-X 2HDM and the characteristics of the Higgs-phobic pseudoscalar. 
  In Sec.~\ref{sec:scanning}, we do the parameter scanning for both old and new sets of $S$ and $T$ values.
   In Sec.~\ref{sec:RGE}, we study the RGE evolutions and the cutoff scales.
   Section \ref{sec:LHC} deals with the LHC phenomenology of the Higgs-phobic type-X. 
   Finally we conclude in Sec.~\ref{sec:conclusions}.

\section{Type-X 2HDM with a Higgs-phobic pseudoscalar boson}
\label{sec:review}

The 2HDM introduces two $SU(2)_L$ complex scalar doublet fields
with hypercharge $Y=+1$, $\Phi_1$ and $\Phi_2$~\cite{Branco:2011iw}:
\bea
\label{eq:phi:fields}
\Phi_i = \left( \begin{array}{c} w_i^+ \\[3pt]
\dfrac{v_i +  \rho_i + i \eta_i }{ \sqrt{2}}
\end{array} \right), \quad (i=1,2)
\eea
where $v_1$ and $v_2$ are the vacuum expectation values  of $\Phi_{1}$ and $\Phi_2$,
respectively.
The ratio of $v_2$ to $v_1$ defines $\tan\bt \equiv v_2/v_1$.\footnote{In what follows,
we will use the simplified notation of
$s_x=\sin x$, $c_x = \cos x$, and $t_x = \tan x$.}
The electroweak symmetry is broken by $v =\sqrt{v_1^2+v_2^2}=246\gev $.
We introduce a discrete $Z_2$ symmetry to prevent the tree-level flavor-changing neutral currents 
(FCNC)~\cite{Glashow:1976nt,Paschos:1976ay},
under which $\Phi_1 \to \Phi_1$ and $\Phi_2 \to -\Phi_2$.
Allowing the softly broken $Z_2$ symmetry and retaining the CP invariance,
we write the scalar potential as
\bea
\label{eq:VH}
V_\Phi = && m^2 _{11} \Phi^\dagger_1 \Phi_1 + m^2 _{22} \Phi^\dagger _2 \Phi_2
-m^2 _{12} ( \Phi^\dagger_1 \Phi_2 + \hc) \\ \nn
&& + \frac{1}{2}\lambda_1 (\Phi^\dagger _1 \Phi_1)^2
+ \frac{1}{2}\lambda_2 (\Phi^\dagger _2 \Phi_2 )^2
+ \lambda_3 (\Phi^\dagger _1 \Phi_1) (\Phi^\dagger _2 \Phi_2)
+ \lambda_4 (\Phi^\dagger_1 \Phi_2 ) (\Phi^\dagger _2 \Phi_1) \\ \nn
&& + \frac{1}{2} \lambda_5
\left[
(\Phi^\dagger _1 \Phi_2 )^2 +  \hc
\right].
\eea

The 2HDM accommodates five physical Higgs bosons, the lighter \textit{CP}-even scalar $h$,
the heavier \textit{CP}-even scalar $H$, the \textit{CP}-odd pseudoscalar $A$,
and a pair of charged Higgs bosons $H^\pm$.
For the relations of the mass eigenstates with the weak eigenstates via two mixing angles of $\al$ and $\bt$, we refer the reader to Ref.~\cite{Song:2019aav}.
The SM Higgs boson $\hsm$ is a linear combination of $h$ and $H$, given by
\bea
\label{eq:hsm}
\hsm = \sba h + \cba H.
\eea
Two scenarios exist in explaining the SM-like Higgs boson~\cite{Aad:2019mbh,CMS:2020xwi,ATLAS:2021vrm},
the normal scenario where $h$ is observed
and the inverted scenario where $H$ is observed while $h$ has been hidden~\cite{Chang:2015goa,Jueid:2021avn,Lee:2022gyf}.
This work focuses on the normal scenario, i.e., $\mh=125\gev$.
Then, the Higgs coupling modifier for a gauge boson pair, $W^+ W^-$ and $ZZ$,
becomes 
\bea
\label{eq:kpV}
\kp_V=\sba.
\eea
If $|\sba|=1$, the couplings of $h$ to the SM particles are the same as in the SM,
which is called the Higgs alignment.

The quartic couplings in \eq{eq:VH} play a crucial role in governing the perturbativity,
unitarity, and vacuum stability.
Near the Higgs alignment limit, 
the quartic couplings are~\cite{Das:2015mwa}
\bea
\label{eq:quartic}
\lm_1 &\simeq& \frac{1}{v^2}
\left[
\mh^2 + \tb^2 \lf \mhh^2 - M^2 \ri
\right],
\\ \nn
\lm_2 &\simeq& \frac{1}{v^2}
\left[
\mh^2 + \frac{1}{\tb^2} \lf \mhh^2 - M^2 \ri
\right],
\\ \nn
\lm_3 &\simeq& \frac{1}{v^2}
\left[
\mh^2 - \mhh^2 -M^2 +2 \mch^2 
\right],
\\ \nn
\lm_4 &\simeq& \frac{1}{v^2}
\left[
M^2+\ma^2-2 \mch^2
\right],
\\ \nn
\lm_5 &\simeq& \frac{1}{v^2}
\left[
M^2-\ma^2
\right],
\eea
where $M^2 = m_{12}^2/(\sb\cb)$.
When $\tb$ is large, the perturbativity of $\lm_1$ is particularly important~\cite{Jueid:2021avn}.
The $\tb^2$ terms in $\lm_1$ easily break the perturbativity and unitarity
unless $M^2$ is almost the same as $\mhh^2$.
The perturbativities of $\lm_4$ and $\lm_5$ with $M^2 \approx \mhh^2$
demand $\mhh$ similar to $\ma$ and $\mch$:
\bea
\label{eq:mass:relation}
M \approx \mhh \sim \ma \sim \mch,
\eea
where $M =\sqrt{M^2}$.

The Yukawa interactions of the SM fermions are parameterized by 
\bea
\label{eq:Lg:Yukawa}
\lg_{\rm Yuk} &=&
- \sum_f 
\lf 
\frac{m_f}{v} \xi^h_f \bar{f} f h + \frac{m_f}{v} \xi_f^H \bar{f} f H
-i \frac{m_f}{v} \xi_f^A \bar{f} \gm_5 f A
\ri
\\ \nn &&
- 
\left\{
\dfrac{\sqrt{2}}{v } \overline{t}
\left(m_t \xi^A_t {P}_- +  m_b \xi^A_b {P}_+ \right)b  H^+
+\sum_{\ell=\mu,\tau}\dfrac{\sqt m_\ell}{v}\xi^A_\ell \,\overline{\nu}_\ell P_+ \ell H^+
+\hc
\right\},
\eea
where $P_\pm=(1\pm\gm^5)/2$.
The Higgs coupling modifiers in type-X are
\begin{align}
\label{eq:xi}
\xi^h_{t,b} &=\sba+\frac{\cba}{\tb},\quad 
\xi^h_\ell = \sba -\cba\tb ,\\ \nn 
\xi^H_{t,b} &=\frac{\sa}{\sb},\quad 
\xi^H_\ell = \frac{\ca}{\cb},\quad
\xi^A_\ell = \frac{1}{\xi^A_t }=  -\frac{1}{\xi^A_b } = \tb.
\end{align}
For the trilinear scalar couplings, we parameterize the Lagrangian as
\bea
\lg_{\rm tri} &=&
v \left[\;
\frac{1}{3!} \sum_{\varphi_0} \lmh_{\varphi_0^3} \varphi_0^3 
+
\frac{1}{2} \lmh_{hhH} \, hh H 
+
\frac{1}{2} \lmh_{hHH} \,h HH
\right.
\\[3pt] \nn && ~~~ \left.
+
 \sum_{\varphi_0} \left\{\frac{1}{2} \lmh_{ \varphi_0AA}\, A^2 \varphi_0
+ \lm_{\varphi_0 H^+ H^-} \,  H^+ H^- \varphi_0\right\}
\right].
\eea
where $\varphi_0=h,H$.

Our central concern is the exotic decay of the observed Higgs boson, $h\to AA$,
which is severely restricted by the current Higgs precision data~\cite{Aad:2019mbh,CMS:2020xwi,ATLAS:2021vrm}.
Since the muon $g-2$ anomaly requires a light pseudoscalar boson
and $h \to AA^* \to A \ttau$ also constrains the model for $\ma > \mhsm/2$,
we need to forbid the $h$-$A$-$A$ vertex.
So we consider type-X with the Higgs-phobic pseudoscalar boson $A$, 
simply called the Higgs-phobic type-X in what follows.
The trilinear coupling for the vertex is
\begin{align}
\lmh_{hAA} & = \frac{1}{4 \sb\cb} 
\left[
\lf 2\ma^2 - \mh^2\ri c_{\al-3\bt} - \lf 2\ma^2 + 3 \mh^2 - 4 M^2 \ri c_{\al+\bt} 
\right].
\end{align}
Since  $\sba$ and $\cba$ are useful when dealing with the Higgs precision data,
we use the identities of
\begin{align}
\frac{c_{\al-3\bt}}{\sb\cb} &= -2 \sba - \lf \tb -\frac{1}{\tb} \ri \cba,
\\[4pt] \nn
\frac{c_{\al+\bt}}{\sb\cb} &=  \phantom{-} 2 \sba - \lf \tb -\frac{1}{\tb}  \ri \cba,
\end{align}
and rewrite $\lmh_{hAA}$ as
\begin{align}
\lmh_{hAA} &= \lf 2 M^2 - 2\ma^2-\mh^2 \ri \sba +(\mh^2 -M^2)\lf \tb -\frac{1}{\tb} \ri \cba.
\end{align}
Then, the condition of $\lmh_{hAA} =0$ accords with 
\bea
\label{eq:Hphobic:condition}
\hbox{Higgs-phobic $A$: } ~~
\frac{\sba}{\cba} = -\lf \tb -\frac{1}{\tb} \ri  \frac{\mhsm^2-M^2}{2 M^2-2\ma^2-\mhsm^2}.
\eea
Note that the exact Higgs alignment cannot coexist with the Higgs-phobic $A$.
Since $\sba$ is determined by $\tb$, $M^2$, and $\ma$,
the model has five parameters of
\bea
\label{eq:parameters}
\{
\tb,\; \ma,\; \mhh,\; \mch,\; M^2
\}.
\eea

\begin{figure}[t] \centering
\begin{center}
\includegraphics[width=0.8\textwidth]{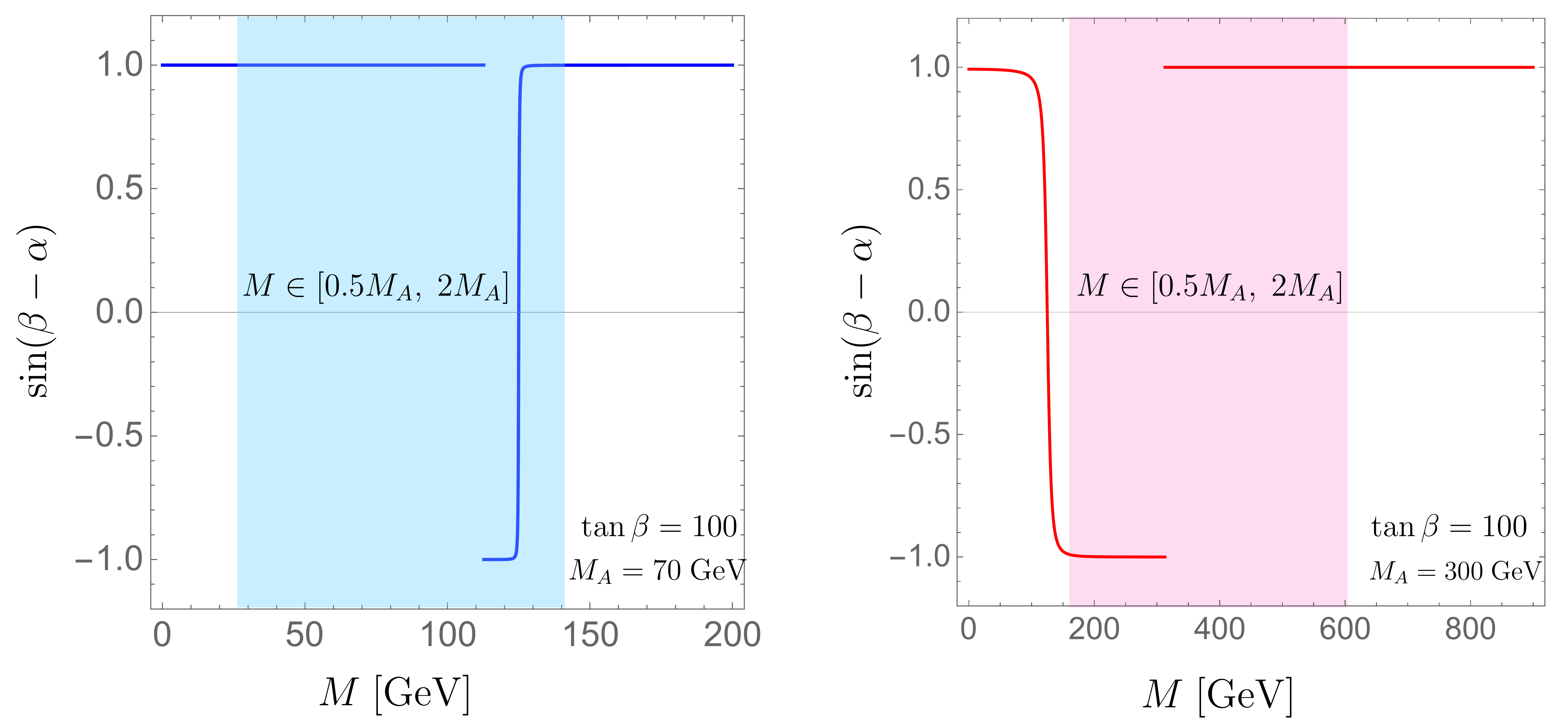}
\end{center}
\caption{\label{fig:Msq-sba-Hphobic}
$\sin (\bt-\al)$ as a function of $M(\equiv \sqrt{M^2})$ in the Higgs-phobic type-X.
For $\tb=100$, we consider $\ma=70\gev$ (left panel) and $\ma=300\gev$ (right panel).
The colored regions correspond to $M \in [0.5\ma,\,2\ma]$.
}
\end{figure}

An interesting consequence of the Higgs-phobic $A$
is that the Higgs alignment naturally arises, although not exact. 
In Fig.~\ref{fig:Msq-sba-Hphobic},
we show $\sba$ as a function of $M$ satisfying \eq{eq:Hphobic:condition}.
Here we take the positive $\cba$ scheme
as in the public codes of \textsc{2HDMC}~\cite{Eriksson:2009ws},
\textsc{HiggsSignals}~\cite{Bechtle:2020uwn}, and \textsc{HiggsBounds}~\cite{Bechtle:2020pkv}.
In Fig.~\ref{fig:Msq-sba-Hphobic},
two cases are considered, $\ma=70\gev$ (left panel) and $\ma=300\gev$ (right panel), with $\tb=100$.
In both cases, $|\sba| \approx 1$ in the most range of $M$.
If we restrict ourselves to $M\sim \ma$,
as shown by the colored regions corresponding to $M \in [0.5\ma,\,2\ma]$,
the preference for the alignment is greater.

Brief comments on the wrong-sign Yukawa coupling of the tau lepton are in order here.
The current Higgs precision data still allow the possibility that $\kp_V$ and $\xi^h_{b,\tau}$ have opposite signs.
In the literature, the LHC phenomenology of the wrong-sign $b$ quark Yukawa coupling in type-II 
has been extensively studied~\cite{Ferreira:2014naa,Coyle:2018ydo,Ferreira:2017bnx,Su:2019ibd,Han:2020zqg}.
In type-X with large $\tb$, however, $\xi^h_{b}$ has the same sign with $\kp_V$.
Wrong-sign Yukawa coupling is only possible for the tau lepton.
In the positive $\cba$ scheme,
caution is needed since the negative sign of $\xi^h_\tau$ does not mean the wrong-sign $\tau$ Yukawa coupling.
If $\sba=-1$, all of the Higgs coupling modifiers have negative sign as $\kp_V=\xi^h_{t,b,\tau}=-1$,
which indicates the right-sign.
In summary, the right-sign and wrong-sign of the tau lepton Yukawa coupling are defined by
\bea
\label{eq:wrong-sign:right-sign}
\hbox{right-sign: }&& \xi^h_\tau \times {\rm sgn}(\sba) >0;
\\ \nn
\hbox{wrong-sign: }&& \xi^h_\tau \times {\rm sgn}(\sba) <0.
\eea

\section{Scanning strategies and the results}
\label{sec:scanning}

Focusing on the Higgs-phobic type-X,
we study the implication of the CDF $m_W$ and muon $g-2$ anomalies
as well as the other theoretical and experimental constraints.
Over the randomly generated parameters in the ranges of
\begin{align}
\label{eq:scan:range}
 t_{\beta} &\in \left[ 1, 200 \right],\quad m_{12}^2 \in \left[ 0, \,15000\right] \gev^2, \\ \nn
M_H &\in \left[ 130,\,1000 \right] \gev, \quad
M_A \in \left[ 10, \, 200\right] \gev, \quad \mch \in [80,\,1000]\gev,
\end{align}
we cumulatively enforce the following constraints in four steps:\footnote{An important constraint is from flavor physics like
$b\to s \gm$~\cite{Arbey:2017gmh,Misiak:2017bgg}.
In type-X, the region with small $\tb$ and the light charged Higgs boson
is significantly constrained:
$\tan \beta > 2.7\,(2.6)$ for $M_{H^+} = 110\, (140)$ GeV~\cite{Arbey:2017gmh}. 
But the observed $\damu$ requires large $\tb$, for which the FCNC processes do not affect.}
\begin{description}
\item[step I:] \textbf{$\damu$\,+\,Theory}
\renewcommand\labelenumi{(\theenumi)}
	\ben
	\item First, we obtain $\sba$ from the model parameters in \eq{eq:parameters}
	by using the Higgs-phobic condition in \eq{eq:Hphobic:condition}.
	For efficient scanning, we preliminary demand $0.8<|\sba|<1$, 
	considering the most updated results on the Higgs coupling modifiers~\cite{Aad:2019mbh}. 
	\item We demand the bounded-from-below potential~\cite{Ivanov:2006yq},
        the unitarity of scalar-scalar scatterings~\cite{Branco:2011iw,Arhrib:2000is},
        the perturbativity of Higgs quartic couplings~\cite{Chang:2015goa}, 
        and the stability of the vacuum~\cite{Ivanov:2008cxa,Barroso:2012mj,Barroso:2013awa}.
        \item We require that the model explains  $\damu$ in \eq{eq:damu:obs}.  
        The contributions to $\damu$ in the 2HDM are summarized in Appendix \ref{appendix:damu}.
       \een
\item[step II:] \textbf{EWPD\,+\,step I}\\ 
We consider the Peskin-Takeuchi oblique parameters $S$ and $T$ with $U=0$ before and after the CDF $m_W$ measurement~\cite{Lu:2022bgw}, called the PDG and CDF cases, respectively:
	\bea
	\label{eq:ST:PDG}
	\hbox{PDG: } && S_{\rm PDG} = 0.05 \pm 0.08, \quad T_{\rm PDG}  = 0.09 \pm 0.07, \quad \rho_{\rm PDG}  = 0.92,\\
	\label{eq:ST:CDF}
	\hbox{CDF: } && S_{\rm CDF}  = 0.15 \pm 0.08, \quad T_{\rm CDF} = 0.27 \pm 0.06, \quad  \rho_{\rm CDF}=0.93, 
		\eea
	where $\rho$ is the correlation between $S$ and $T$. 
	In the 2HDM, the oblique parameters  have been extensively studied~\cite{Toussaint:1978zm,Bertolini:1985ia,Pomarol:1993mu,Peskin:2001rw,He:2001tp,Grimus:2007if,Grimus:2008nb,Kanemura:2011sj}.
	We use the public code \textsc{2HDMC}~\cite{Eriksson:2009ws},
	which adopt the calculation of Refs.~\cite{Grimus:2007if,Grimus:2008nb}.
	Then, we perform the $\chi^2$ analysis in the $(S,T)$ plane, requiring $p>0.05$.
\item[step III:] \textbf{Collider\,+\,step II}
\renewcommand\labelenumi{(\theenumi)}
	\ben
	\item The Higgs precision data are checked via the public code \textsc{HiggsSignals}-v2.6.2~\cite{Bechtle:2020uwn}
	which takes into account 111 Higgs observables~\cite{Aaboud:2018gay,Aaboud:2018jqu,Aaboud:2018pen,Aad:2020mkp,Sirunyan:2018mvw,Sirunyan:2018hbu,CMS:2019chr,CMS:2019kqw}.
	Since our model has five parameters,  the number of degrees of freedom is 106.
Based on 	the $\chi^2$ value from the \textsc{HiggsSignals},
we demand that the $p$-value should be larger than 0.05.
	\item The direct searches for BSM Higgs bosons
	at the LEP, Tevatron, and LHC are examined by using the open code 
	\textsc{HiggsBounds}-v5.10.2~\cite{Bechtle:2020pkv}.
We exclude a parameter point
if any cross section predicted by the model exceeds the observed 95\% C.L. upper bound.
	\een	
\item[step IV:] \textbf{LFU\,+\,step III}
\\
We perform a global $\chi^2$ fit of the Higgs-phobic type-X to $\damu$ and the following LFU data:
	\ben
	\item For the $\tau$ decay, we adopt the HFLAV global fit results of~\cite{HFLAV:2019otj}
	\bea
	\label{eq:HFLAV}
	\frac{g_\tau }{ g_\mu}, \quad
 \frac{g_\tau}{ g_e}  , 
 \quad
\frac{g_\mu}{g_e}  , 
 \quad
\lf \frac{g_\tau}{ g_\mu }\ri_\pi, 
 \quad
\lf g_\tau \over g_\mu \ri_K.
	\eea
	One redundant degree of freedom should be removed since it has 
	 a zero eigenvalue in the covariance matrix.
	\item 
	We include the Michel parameters~\cite{Michel:1949qe,Logan:2009uf}
	from  the energy and angular distributions of $\ell^-$ in the decay of $\tau^- \to \ell^- \nu\nu_\tau$:
	\begin{eqnarray}
	\label{eq:Michel:def}
 \rho_e , \quad 
 \lf \xi \delta \ri_e , 
\quad 
 \xi_e ,  \quad
\eta_\mu , 
\quad 
\rho_\mu, 
\quad 
\lf \xi \delta \ri_\mu , 
\quad 
 \xi_\mu , 
\quad  \xi_\pi , 
\quad
\xi_\rho , 
\quad
\xi_{a_1} .
\end{eqnarray}
	\item We also incude the accurate measurement of the leptonic $Z$ decays.
	Two ratios of the partial decay rates are considered~\cite{ALEPH:2005ab}:
\bea
\label{eq:obs:Z}
\frac{\Gm(Z\to \mmu)}{\Gm(Z \to \ee)}, \quad
\frac{\Gm(Z\to \ttau)}{\Gm(Z \to \ee)}.
\eea
	\een
The theoretical calculations of the LFU observable in type-X are summarized in Appendix \ref{appendix:LFU}
and the experimental data are referred to Ref.~\cite{Jueid:2021avn}.
Including $\damu$,
we have 17 independent observables in the global fit.
Since the model parameters have already been restricted through step I, II, and III,
we consider the number of degrees of freedom to be $N_{\rm dof}=17$
and demand the $p$-value larger than 0.01.
In the SM, the $p$-value is only $0.003$~\cite{Jueid:2021avn}.

	\end{description}
	
We randomly scan the five-dimensional parameter space in \eq{eq:scan:range}.
For the PDG and CDF cases,
we independently obtained $10^7$ parameter points that pass step I.
Setting step I as the reference, we calculate the survival probabilities at each step:
\bea
\label{eq:survival:probabilities}
\hbox{PDG: } && P_{\rm step II} = 5.47\%, \quad P_{\rm step III} = 3.15\%, \quad P_{\rm step IV} = 0.62\%, 
\\ \nn
\hbox{CDF: } && P_{\rm step II} = 1.56\%, \quad P_{\rm step III} = 1.00\%, \quad P_{\rm step IV} = 0.21\%.
\eea
The Higgs-phobic type-X does have considerable parameter points that explain all the 
constraints.
The validity of the model is largely irrelevant to whether we take the PDG or CDF case, but
the survival probabilities are different.
The PDG case has approximately three times greater probability than the CDF.
But just because the PDG case has more viable parameter points does not mean 
it is a better solution.
%

 \begin{figure}[h] \centering
\begin{center}
\includegraphics[width=0.45\textwidth]{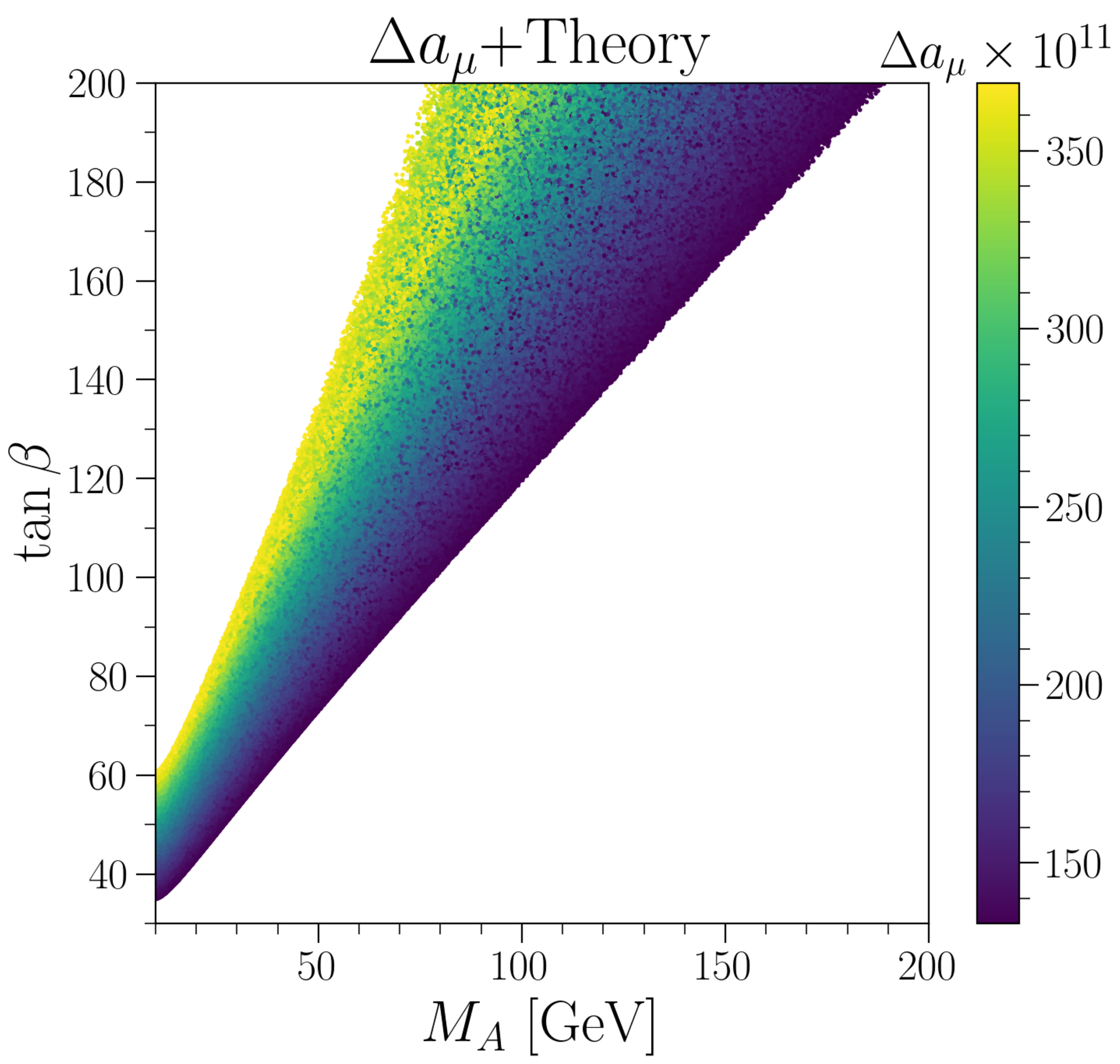}
\end{center}
\caption{\label{fig:step I-tb-MA-damu}
Allowed regions of $(\ma,\tan\beta)$ at step I with $\damu$ and the theoretical
constraints.
The color code indicates $\damu$.}
\end{figure}

Now we investigate which constraint excludes which region of the parameter space.
First, we present $\tb$ versus $\ma$ at step I in Fig.~\ref{fig:step I-tb-MA-damu},
which is common for the PDG and CDF cases.
The color code indicates $\damu$.
The observed $\damu$ allows the band shape in $(\ma,\,\tb)$.
We need large $\tb $ above $\sim 35$
and light $\ma$ below $\sim 170\gev$.
$\ma$ above 170 GeV is also feasible if $\tb$ is greater than 200.
But we avoid too large $\tb$ to retain 
the perturbativity of the Yukawa coupling of the tau lepton to the BSM Higgs bosons.

 \begin{figure}[h] \centering
\begin{center}
\includegraphics[width=0.98\textwidth]{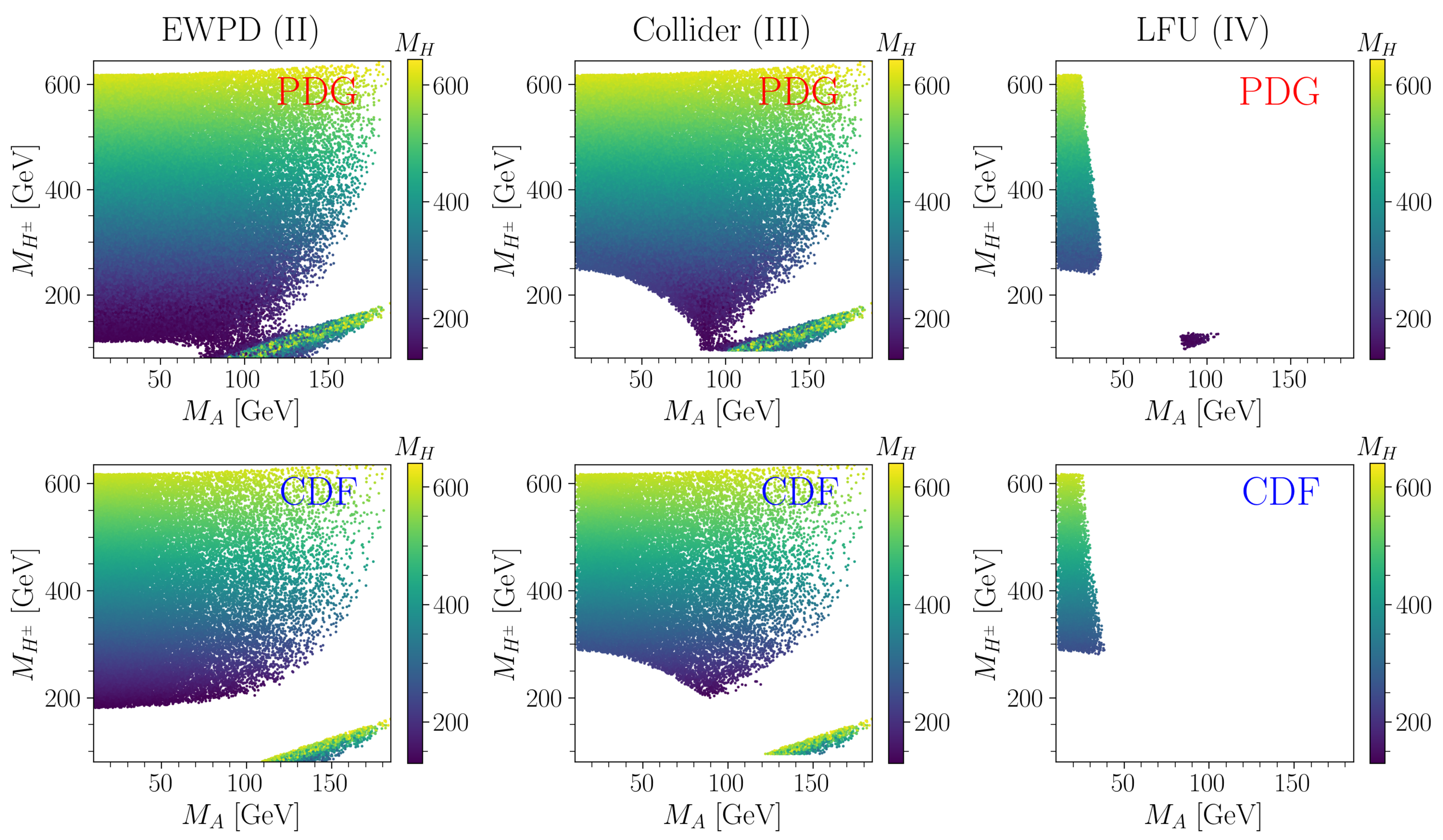}
\end{center}
\caption{\label{fig:McH-MA-MH}
$\mch$  versus $\ma$ at step II (left panels), step III (middle panels), and step IV (right panels),
with the color code indicating $\mhh$.
We consider the PDG case (upper panels) and the CDF case (lower panels).}
\end{figure}

As we go through the remaining steps,
the masses of the other BSM Higgs bosons are also constrained.
In Fig.~\ref{fig:McH-MA-MH},
we show $\mch$ versus $\ma$ with the color code of $\mhh$
at step II (left panels), step III (middle panels), and step IV (right panels).
We compare the PDG case (upper panels) with the CDF (lower panels). 
Let us begin with their common features.
The first and most important one is that upper bounds exist on
 the masses of new Higgs bosons, which appear in step II.
It is because the light $\ma$, which is required to explain $\damu$,
brings down $\mhh$ and $\mch$ to yield small $S$ and $T$.
The upper bounds on $\mhh$ and $\mch$ remain almost intact to the last step
such that $M_{H,\ch}\lsim 600\gev$ in both cases.

 \begin{figure}[h] \centering
\begin{center}
\includegraphics[width=0.9\textwidth]{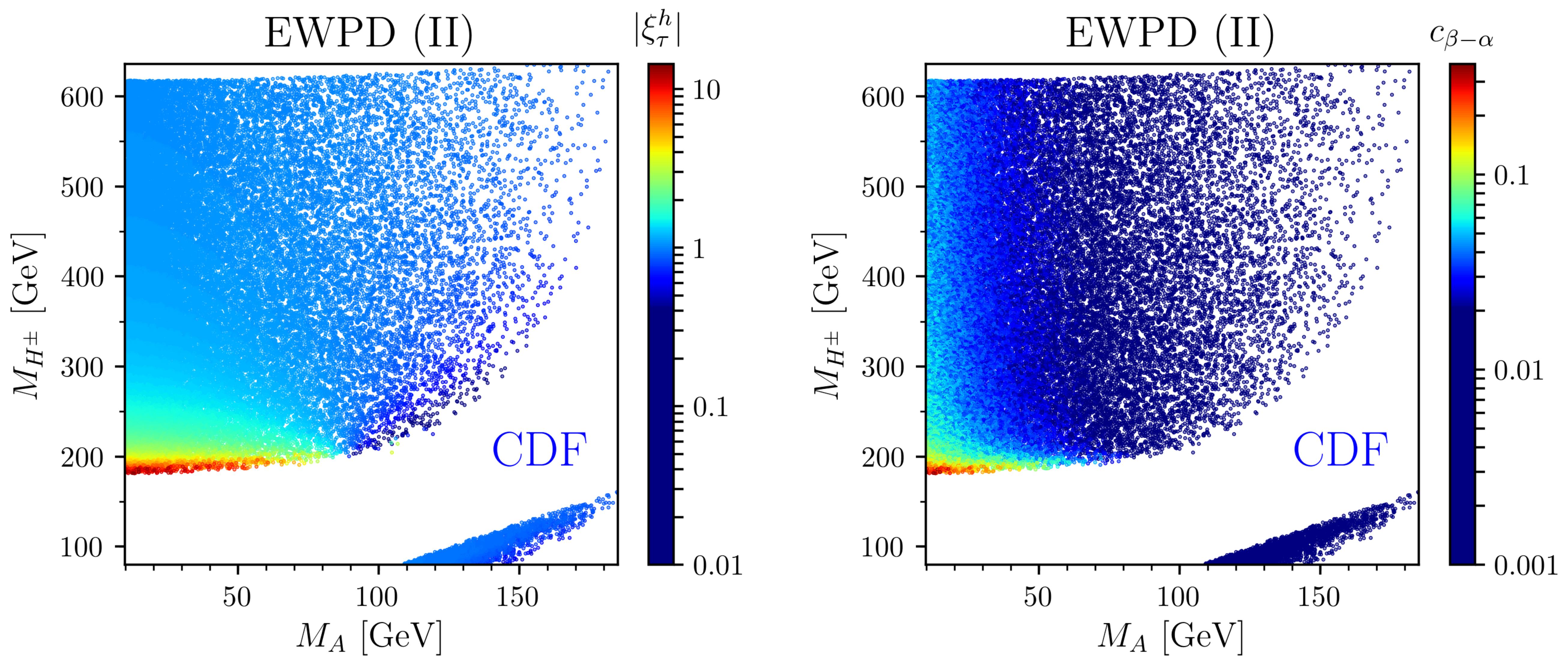}
\end{center}
\caption{\label{fig:xihtau:cba}
For the parameter points that pass step II,
$\mch$ versus $\ma$ with the color code of $|\xi^h_\tau|$ (left panel) and 
with the color code of $\cba$ (right panel).
We focus on the CDF case.}
\end{figure}

The second common feature is the exclusion of the lower-left corner in $(\ma,\mch)$ at step III (Collider), 
mainly from $h\to \ttau$.
In Fig.~\ref{fig:xihtau:cba},
we show for the CDF case 
$\mch$ versus $\ma$ with the color code of $|\xi^h_\tau |$ (left panel) and
$\cba$ (right panel)
over the parameter points that pass step II (EWPD).
As can be seen in the left panel of Fig.~\ref{fig:xihtau:cba},
the area that disappears as we go from step II to step III in Fig.~\ref{fig:McH-MA-MH}
almost coincides with the region of too large $|\xi^h_\tau|$.
This behavior is attributed to $\xi^h_\tau$ in \eq{eq:xi}.
When the Higgs alignment is broken even a little, large $\tb$ increases $|\xi^h_\tau|$ unacceptably.
To reveal the feature in more detail, 
we present $\cba$ via the color code over the plane of $(\ma,\mch)$ in the right panel of Fig.~\ref{fig:xihtau:cba}.
The region with light $\ma$ and light $\mch$ has relatively sizable $\cba$,
which further enhances $|\xi^h_\tau|$.
So, the exclusion by $h\to \ttau$ results in the lower bound on $\mch$ for light $\ma$.

The third common feature is that the global fit to $\damu$ and the LFU data
removes most of the parameter space with $\ma \gsim 38\gev$:
the exceptional island-shaped region in the PDG case
is deferred until we discuss the differences between the PDG and CDF cases.
The exclusion of $\ma\gsim 38\gev$ is primarily from the tree-level contributions to the lepton flavor violating decays of 
the tau lepton, 
mediated by the charged Higgs boson.
The key parameter is~\cite{Jueid:2021avn}
\bea
\label{eq:dt:tree:main}
\dt_{\rm tree} = \frac{m_\mu m_\tau \tb^2}{\mch^2}.
\eea 
Large $\tb$, which corresponds to heavy $\ma$ because of $\damu$, blows up the $\chi^2_{\rm LFU}$ value.
So only the region with very light $\ma$ is finally allowed.

Even though the PDG and CDF cases share many common features,
significant differences also exist.
The first noticeable difference is the island-shaped region at step IV in the PDG case.
To facilitate discussion below, let us call this special region the PDG-island and call the bulk region with $\ma\lsim 38\gev$
the mainland.
The parameters in the PDG-island are populated around
\begin{align}
\label{eq:PDG-island}
\hbox{PDG-island: } & \mhh  \in  [130.0, 165.3] \gev,\quad
\ma\in [84.1, 111.9]\gev, \\ \nn
& \mch \in  [96.5, 127.9]\gev, \quad
\tb > 154.9.
\end{align}
In the CDF case, however, the parameter points in \eq{eq:PDG-island} are excluded from step II.
To understand the origin, let us present the oblique parameter $T$ in the limit of $ \ma\simeq\mhh\simeq\mch$:
\bea
T \simeq \frac{\dma\dmhh}{12 \pi^2 \al \, v^2 }, \quad \hbox{ if } \mch\simeq\ma\simeq\mhh,
\eea
where $\Dt M_i = M_i - \mch$.
The $T_{\rm CDF}$ in \eq{eq:ST:CDF} requires $\Dt M_{A,H} \gsim 80\gev$
that the PDG-island cannot satisfy.
On the contrary,
$T_{\rm PDG}$ permits the mass degeneracy among BSM Higgs bosons, which the PDG-island requires.

 \begin{figure}[t] \centering
\begin{center}
\includegraphics[width=0.5\textwidth]{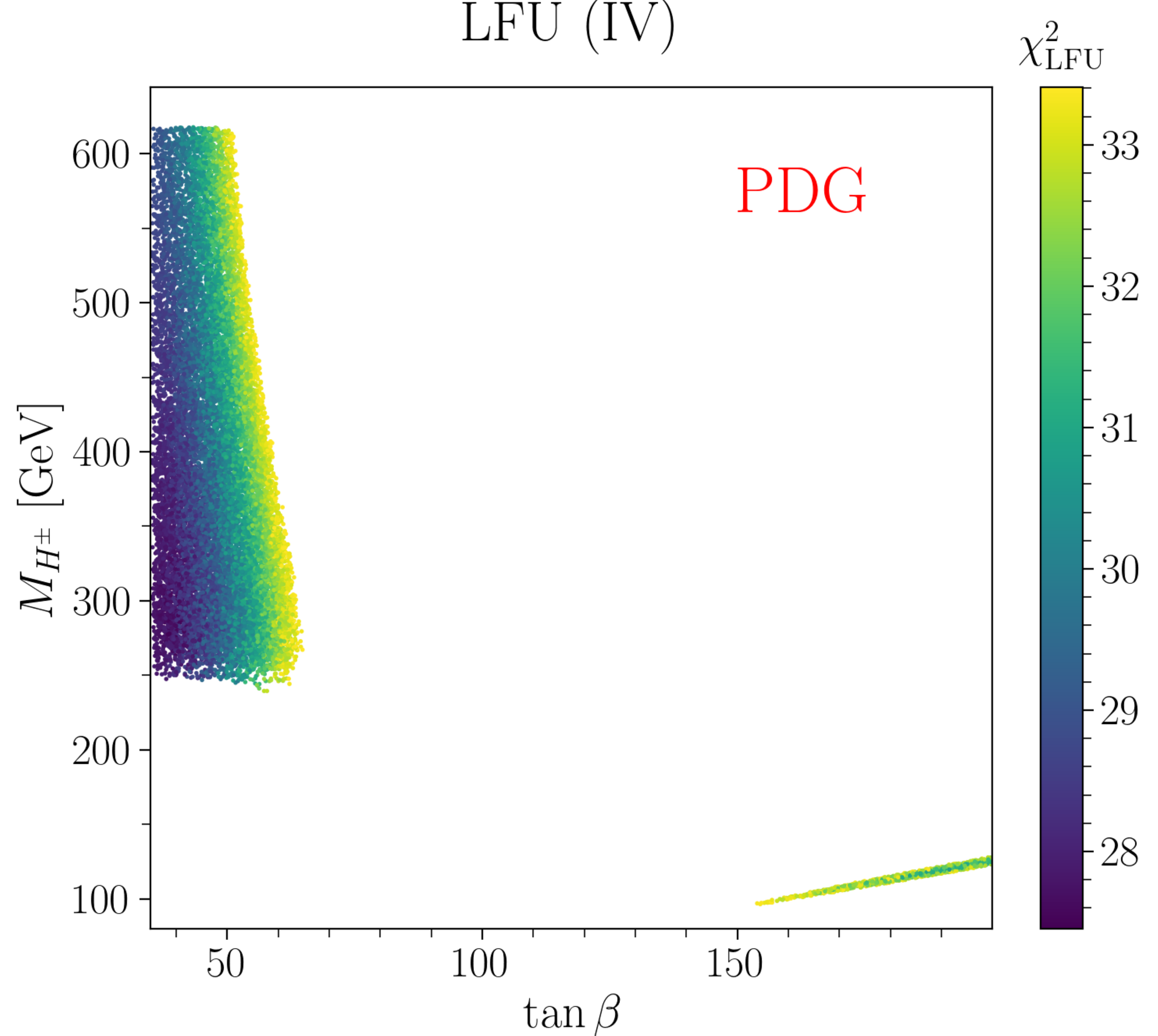}
\end{center}
\caption{\label{fig:McH-tb-chi2}
$\mch$  versus $\tb$ at step IV
with the color code indicating $\chi^2_{\rm LFU}$.
We focus on the PDG case.}
\end{figure}

An important question about the PDG-island is how it can evade the most profound constraints from the LFU data.
As discussed before, the key parameter $\dt_{\rm tree}$ in \eq{eq:dt:tree:main} requires small $\tb$ and thus light $\ma$.
But there exists an alternative way to evade the LFU constraints
through another key parameter of
\bea
\label{eq:es:tree:main}
\es^\tau_{\rm tree} &=& \dt_{\rm tree} 
\left[\frac{ \dt_{\rm tree}}{8}
-\frac{m_\mu}{m_\tau} \frac{g\lf \rho^\mu_\tau \ri}{f\lf \rho^\mu_\tau \ri}
\right],
\eea 
where $g(x)$, $f(x)$, and $\rho^i_j$ are referred to Appendix~\ref{appendix:LFU}.
If the first and second terms in \eq{eq:es:tree:main} are exquisitely cancelled,
the value of $\chi^2_{\rm LFU}$ can be substantially reduced.
The cancellation demands a relation of $\mch$ to $\tb$.
In Fig.~\ref{fig:McH-tb-chi2},
we show $\mch$ versus $\tb$ with the color code of $\chi^2_{\rm LFU}$
over the finally allowed parameter points in the PDG case.
Here we only show the parameter points with $\chi^2_{\rm LFU} < 33.41$, i.e., $p>0.01$ with 17 degrees of freedom.
It is clearly seen that the minimum of $\chi^2_{\rm LFU}$
occurs in the mainland region with $\mch\gsim 250\gev$ and $\tb\simeq 35$.
Almost all the parameter points outside the mainland
have $p$-value below 0.01.
Exceptional is the band-shape PDG-island with $\mch\in  [96.5, 127.9]\gev$ and $\tb >154.9$,
which accommodates the cancellation in \eq{eq:es:tree:main}.

The second difference between the PDG and CDF cases is the lower bound on $\mch$ for $\ma\lsim 38\gev$:
$\mch\gsim 250\gev$ in the PDG case while $\mch\gsim 300\gev$ in the CDF case.
The difference begins in step II.
When $\ma\ll \mch$,
$S$ and $T$ are approximated into
\begin{align}
\label{eq:T}
S &\simeq -\frac{5}{72\pi},
\\[4pt] \nn
T &\simeq -\frac{\mch \Dt \mhh}{16\pi^2 \al\, v^2}
\left[ 1- \frac{\Dt \mhh}{6 \mch}  + \mco \lf \frac{\Dt \mhh^3}{\mch^3} \ri \right].
\end{align}
The positive $T_{\rm CDF}$ in \eq{eq:ST:CDF} prefers negative and nonzero $\Dt \mhh $ for light $\ma$.
Therefore, the heavy mass of $H$, above $125\gev$ by definition, pushes up
the lower bound on $\mch$ in the CDF case.
The substantial mass gap between $\ma$ and $\mch$ guarantees the dominant decay mode of $\ch\to \wpm A$.
In the PDG case, there are two different regions in the charged Higgs boson phenomenology,
the mainland region with $\mch\gsim 250\gev$ 
and the island region with $\mch\simeq 100\gev$.

\begin{figure}[h] \centering
\begin{center}
\includegraphics[width=0.98\textwidth]{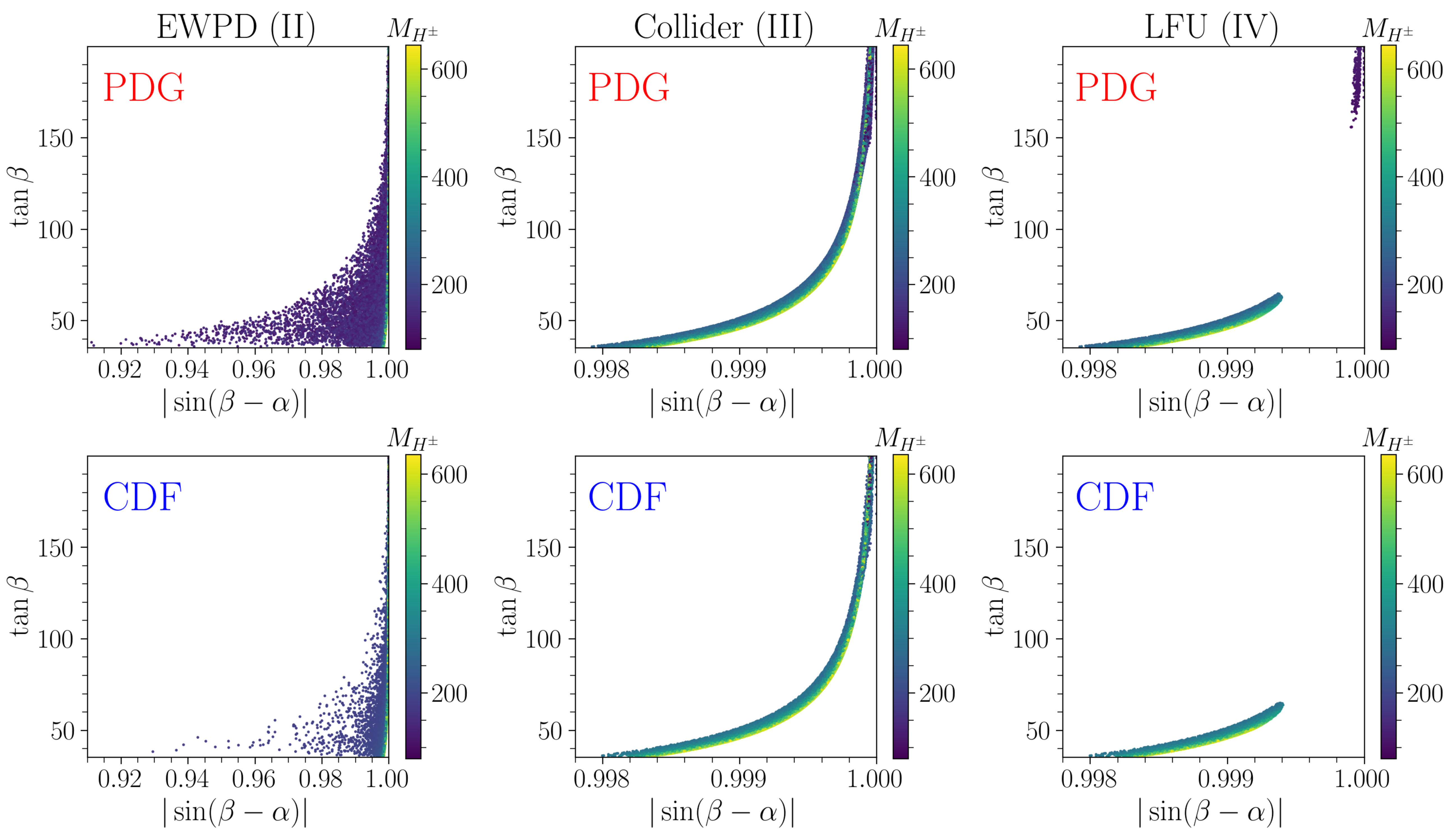}
\end{center}
\caption{\label{fig:tb-sba-McH}
$\tan\beta$ versus $|\sin (\bt-\al)|$ with color code of $\mch$ at step II (left panels), step III (middle panels), and step IV (right panels).
We compare the PDG (upper panels) and the CDF (lower panels).}
\end{figure}

The third difference 
is found in the allowed $\tb$ and $\sba$.
In Fig.~\ref{fig:tb-sba-McH},
we present $\tb$ versus $|\sba|$ with the color code of $\mch$
at step II (left panels), step III (middle panels), and step IV (right panels).
We compare the results of the PDG (upper panels) with those of the CDF (lower panels).
The generic feature of the Higgs-phobic type-X, 
the almost exact Higgs alignment, appears from step II.
When imposing the Higgs precision data at step III,
the tendency toward the Higgs alignment is stronger.
A dramatic change occurs in step IV.
Large $\tb$ above $\sim 65$ is excluded in the CDF case.
In the PDG case, however,
the region with $\tb\in [170,\,200]$ and $|\sba| \approx 1$ remains,
corresponding to the PDG-island.

\begin{figure}[h] \centering
\begin{center}
\includegraphics[width=0.5\textwidth]{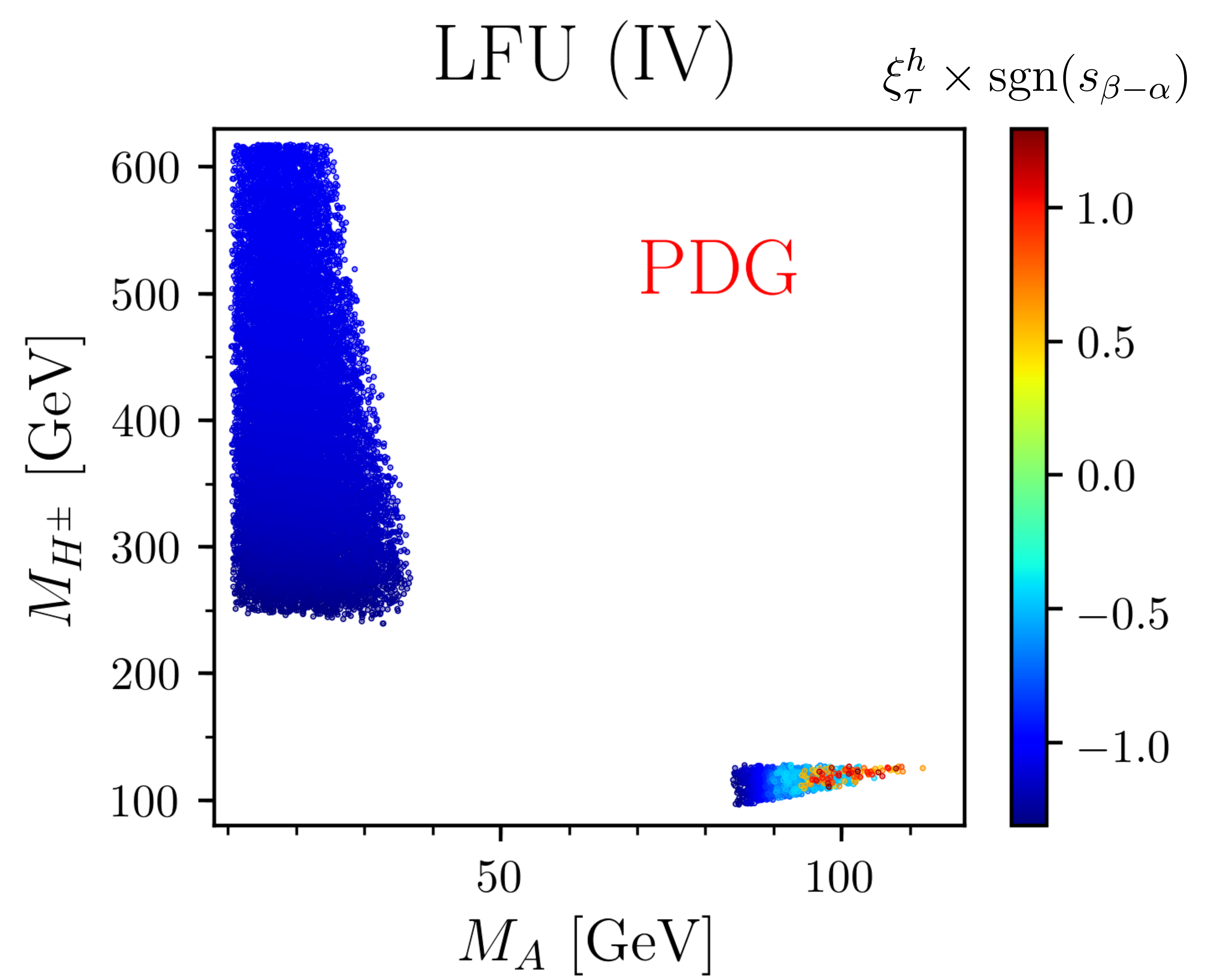}
\end{center}
\caption{\label{fig:wrong-sign}
$\mch$ versus $\ma$ with the color code of $\xi^h_\tau \times {\rm sgn}(\sba)$ in the PDG case.
}
\end{figure}

The last difference is the sign of the tau lepton Yukawa coupling.
Considering the definitions of the right-sign and wrong-sing $\tau$ Yukawa coupling in \eq{eq:wrong-sign:right-sign},
we present $\xi^h_\tau \times {\rm sgn}(\sba)$ via color codes
over the parameter space of $(\ma,\mch)$ in Fig.~\ref{fig:wrong-sign}.
The mainland with $\ma\lsim 38\gev$, in the PDG and CDF cases, has wrong-sign $\tau$ Yukawa coupling,
as discussed in Ref.~\cite{Chun:2015hsa}.
In the PDG-island, however,
right-sign $\tau$ Yukawa coupling is also possible in a sizable portion, about 10\%, of the finally allowed parameter space.
It is attributed to almost 100\% alignment in the PDG-island (see Fig.~\ref{fig:wrong-sign}):
if $\cba$ is small enough to suppress the large $\tb$ in \eq{eq:xi},
$\xi^h_\tau$ and $\sba$ have the same sign.
Probing the wrong-sign $\tau$ Yukawa coupling at the LHC
will give us an important implication on the PDG-island.

\section{Cutoff scales via the RGE analysis}
\label{sec:RGE}

Now that the Higgs-phobic type-X is shown to explain all the constraints, 
a question arises as to what energy scale this model is valid.
To answer the question, we run each parameter point via the RGE
and check three conditions---unitarity, perturbativity, and vacuum stability---as increasing the energy scale.
If any condition is broken at a particular energy scale,
we stop the evolution and record the energy scale as the cutoff scale $\lmc$.

We use the public code \textsc{2HDME}~\cite{Oredsson:2018yho,Oredsson:2018vio}
to run the following parameters:
\bea
\label{eq:running:parameters}
g_s,\quad g,\quad g', \quad \lm_{1,\cdots,5}, \quad \xi^{h,H,A}_f,\quad m_{ij}^2, \quad v_{i}, \quad (i=1,2).
\eea
First, we convert the model parameters in \eq{eq:parameters}
into those in \eq{eq:running:parameters}.
The top quark pole mass of $m_{t}^{\rm pole} = 173.4\gev$
is used to match the 2HDM to the SM parameters.
The boundary conditions at $m_{t}^{\rm pole}$ are referred to Ref.~\cite{Oredsson:2018yho}.
And we evolve them into higher energy scale through
the one-loop RGE.\footnote{The two-loop results are not substantially different from the one-loop results.}

\begin{figure}[t!] \centering
\begin{center}
\includegraphics[width=0.9\textwidth]{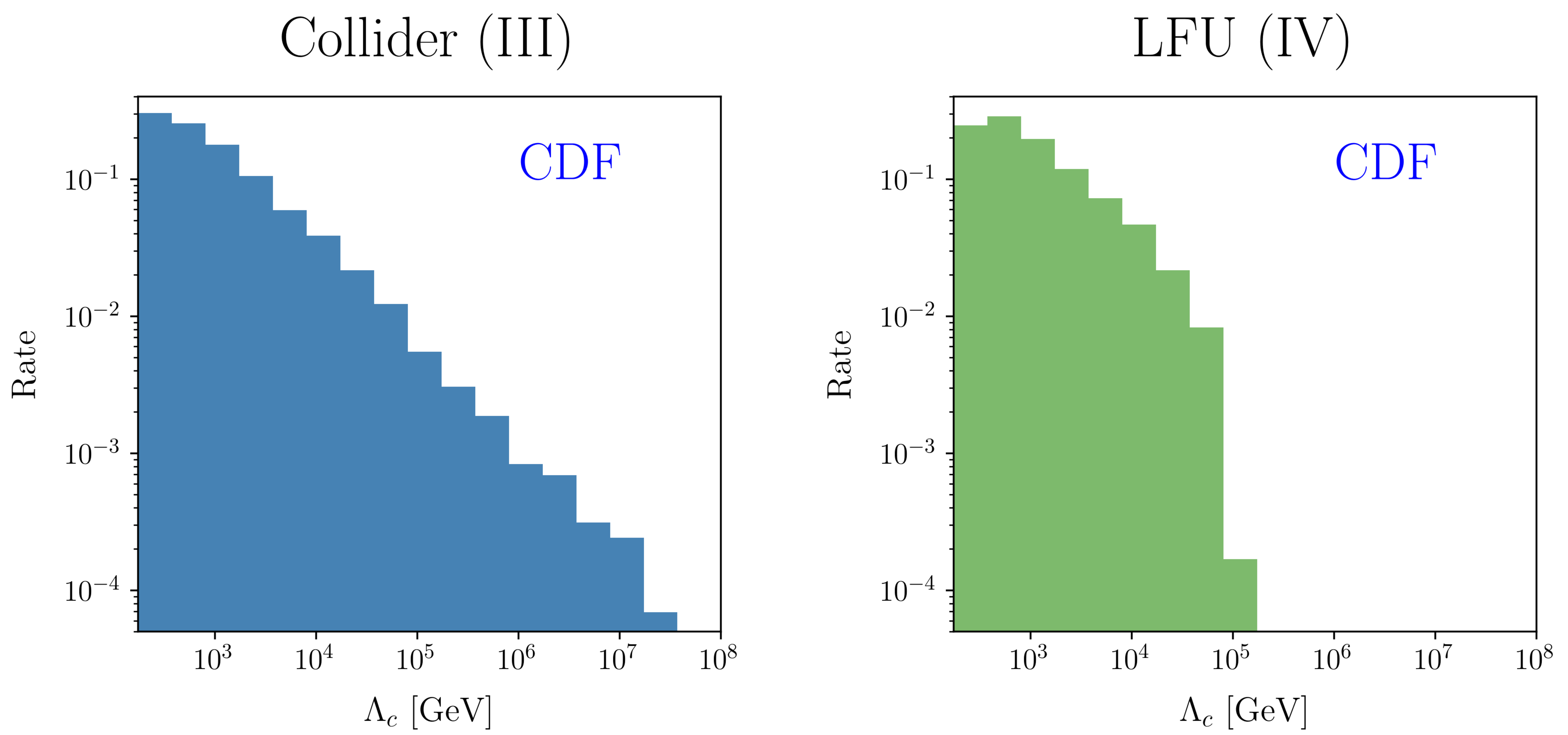}
\end{center}
\caption{\label{fig:RGE:CDF}
Distributions of the cutoff scales of the parameter points at step III (left panel)
and step IV (right panel) in the CDF case.}
\end{figure}

To present the high energy scale behavior of all the viable parameter points, we show 
the distribution of $\lmc$ in Fig.~\ref{fig:RGE:CDF}, focusing on the CDF case.
We compare the $\lmc$ distribution of the parameter points at step III (left panel) with 
those at step IV (right panel).
The \enquote{Rate} in the $y$-axis denotes the ratio $N_\lmc/N_{\rm step}$,
where  $N_\lmc$ is the number of the parameter points with the cutoff scale $\lmc$ and 
$N_{\rm step}$ is the total number of the parameter points at step III (left panel) and at step IV (right panel).
At step III, the Higgs-phobic type-X is stable up to about $10^7\gev$.
After step IV, however,
the model is valid only up to about $10^5\gev$.
Although the Higgs-phobic type-X is a viable model at the electroweak scale,
it needs an extension at the energy scale not far from the LHC reach.
Future colliders at $\sqrt{s}=100\tev$,
such as  the Future hadron-hadron Circular Collider (FCC-hh) at CERN  \cite{Gomez-Ceballos:2013zzn}
and the CEPC~\cite{Gao:2021bam, CEPCStudyGroup:2018ghi},
are expected to find a hint of the next-level BSM model.

If we further require a high cutoff scale,
the parameter space is considerably constrained.
For $\lmc>1\tev$,
the surviving probability is almost halved.
If $\lmc>10\tev$,
the survival probability in the CDF case goes down to 0.01\%
with the parameter points of
\bea
\hbox{if }\Lm_{\rm c}^{\rm CDF}>10\tev:&& \ma \in   [11, \, 38] \gev,\quad 
\mhh \in  [249, \, 306] \gev,\quad 
\\ \nn
&&
\mch \in  [283, \, 338] \gev,\quad  M \in  [249,\, 306] \gev,
\\ \nn 
&&
\tb \in [36.6, \,64.7].
\eea
Since the BSM Higgs boson masses are within the LHC reach,
we expect that the HL-LHC can probe the model with high $\lmc$.

\begin{figure}[t!] \centering
\begin{center}
\includegraphics[width=0.9\textwidth]{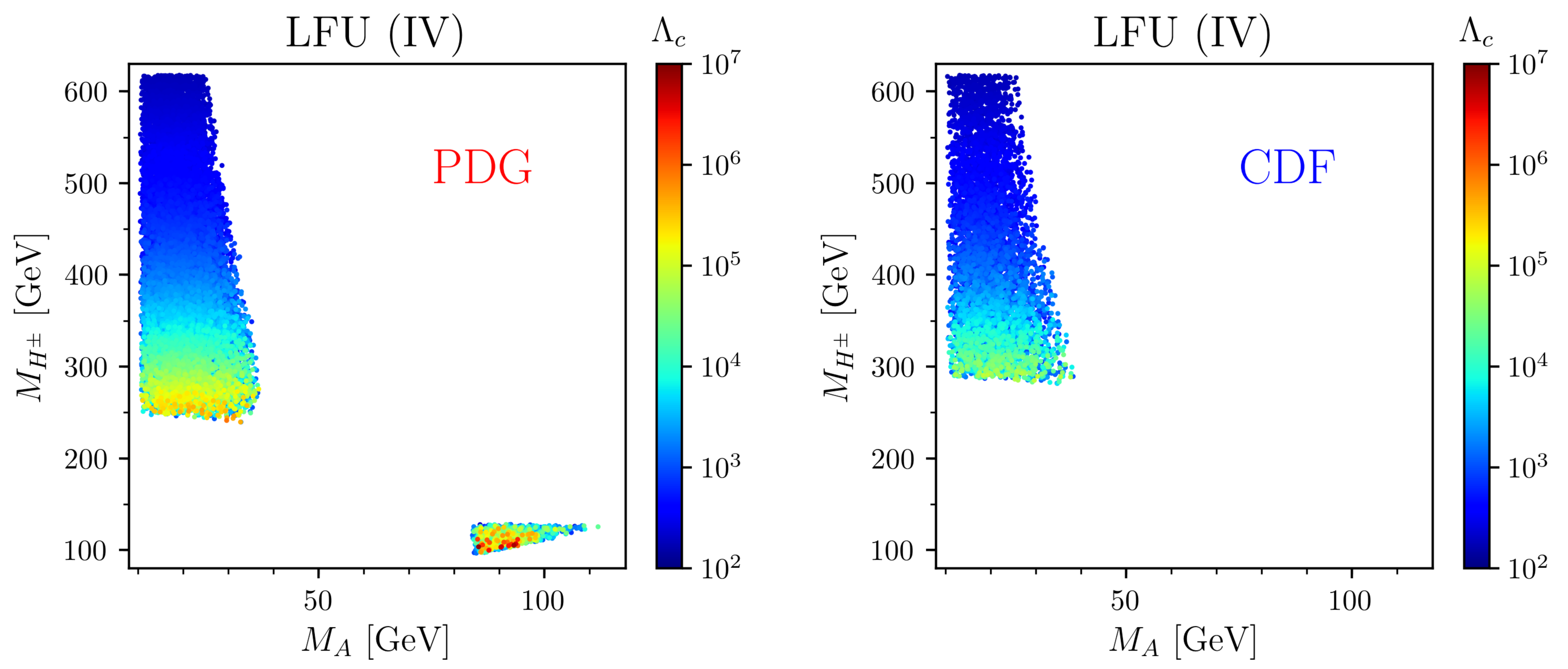}
\end{center}
\caption{\label{fig:MA-McH-cutoff}
Cutoff scales via the color code
in the finally allowed $(\ma,\mch)$.
The left (right) panel shows the results in the PDG (CDF) case.}
\end{figure}

The final discussion is on the difference in the high-energy scale behaviors between the PDG and CDF cases.
In Fig.~\ref{fig:MA-McH-cutoff},
we present the cutoff scales via the color code
in the finally allowed $(\ma,\mch)$.
The left (right) panel shows the results in the PDG (CDF) case.
The difference is clear.
The PDG case can accommodate a larger cutoff scale.
In the mainland region with $\ma \lsim 38\gev$,  $\lmc$ can go up to $10^6\gev$,
which is about ten times higher than $\lmc$ in the CDF case.
In the PDG-island, the cutoff scale is much higher up to about $10^7\gev$.
In terms of the high energy scale stability,
the PDG-island is the most attractive.

\section{Golden discovery channels at the LHC}
\label{sec:LHC}

For the LHC phenomenology of the Higgs-phobic type-X, 
we first study the branching ratios of the BSM Higgs bosons.
The pseudoscalar boson decays only into the fermionic sector:
neither light $\ma \lf \lsim 38\gev \ri$ nor approximately degenerate $\ma$ with $M_{H,\ch}$ in the PDG-island
can accommodate the bosonic decays of $A \to \ch W^{\pm (*)} / H Z^{(*)}$.
Furthermore, the suppressed couplings of $A$ to the quark sector by large $\tb$
make $A\to \tau^+\tau^-$ dominant~\cite{Kanemura:2011kx,Chun:2018vsn}: its branching ratio is almost 100\%. 
Another interesting decay channel is $A\to \mmu$.
Although it has a small branching ratio of about 0.3\%,
the absence of neutrinos helps reconstruct the pseudoscalar mass.
On the other hand, 
$H^\pm$ and $H$ can have the bosonic decay modes of $H^\pm \to W^\pm A$ and $H\to ZA$ for light $\ma$.
Since their partial decay widths 
are enhanced by a factor of $(\mch^2/m_{W}^2)^2$ and $(\mhh^2/m_{Z}^2)^2$ respectively,
$H^\pm \to W^\pm A$ and $H\to ZA$ are dominant 
in the mainland regions.

\begin{figure}[t] \centering
\begin{center}
\includegraphics[width=0.8\textwidth]{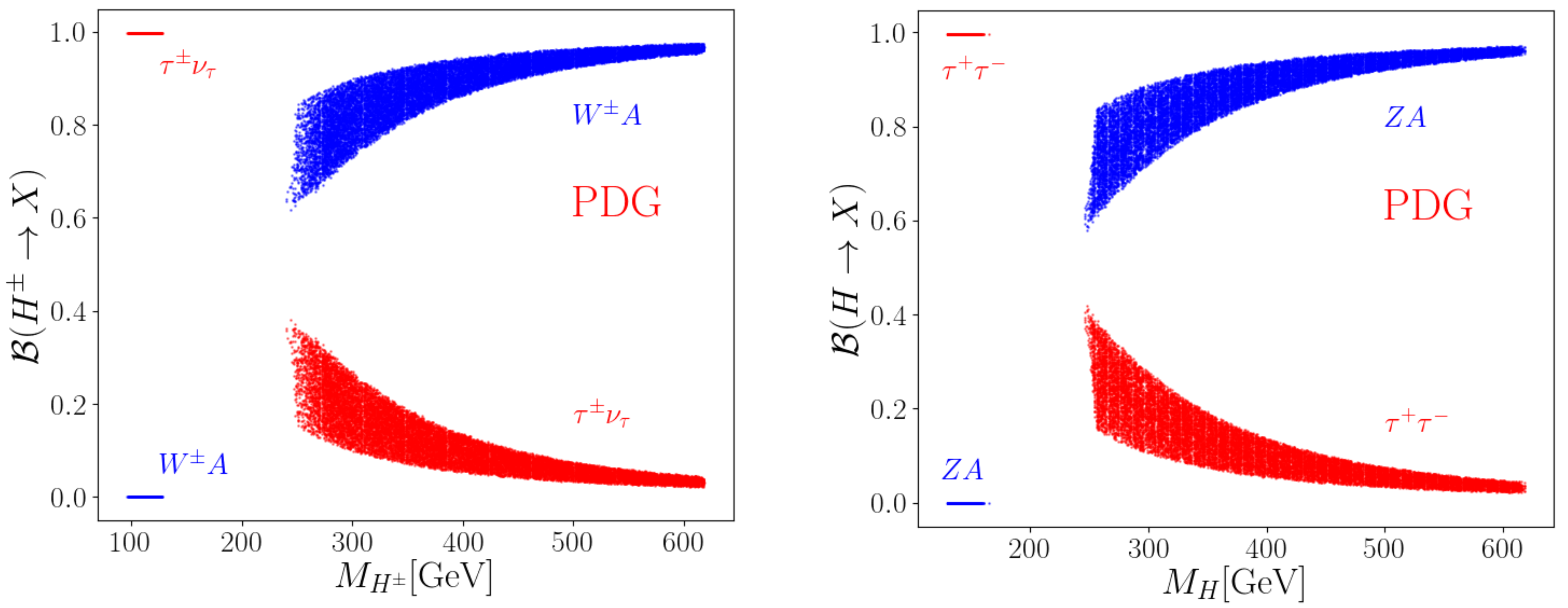}\\[3pt]
\includegraphics[width=0.8\textwidth]{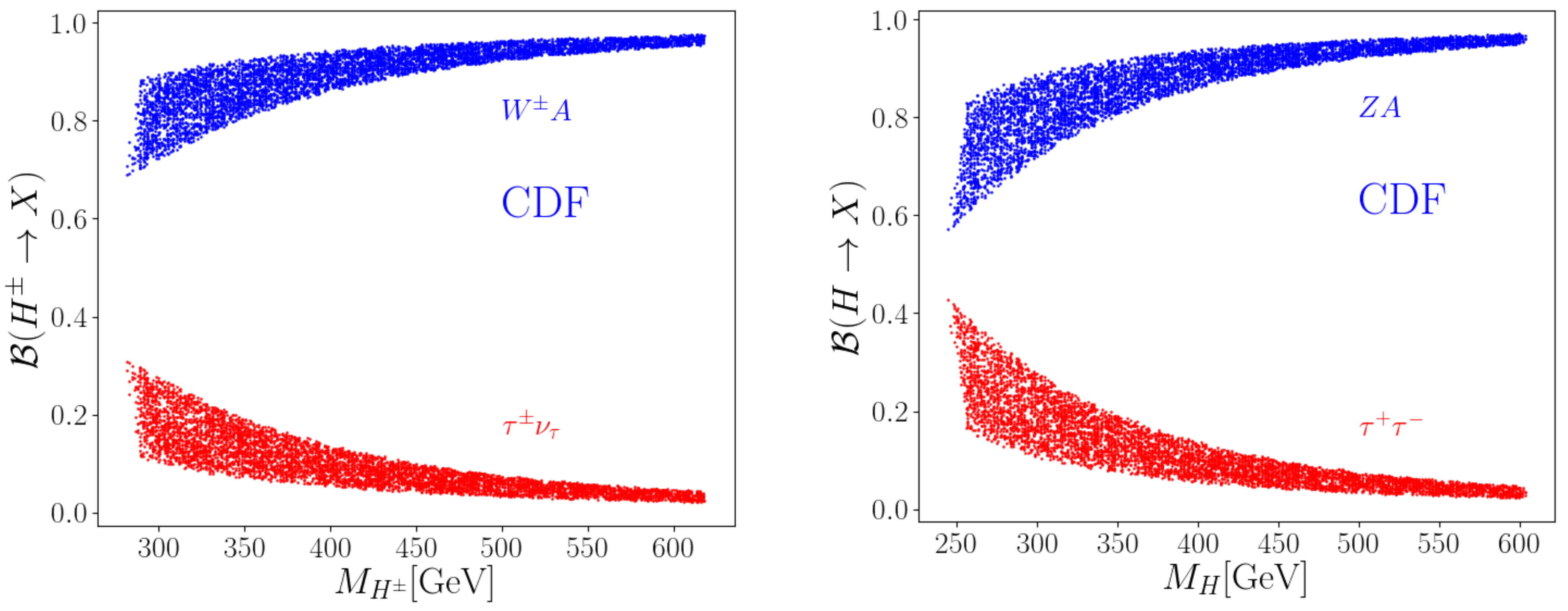}
\end{center}
\caption{\label{fig:Branching_Ratio}
Branching ratios of $\ch$ (left panels) and $H$ (right panels) in the PDG (upper panels) and CDF cases (lower panels),
over the parameter points at the final step IV.
The muon modes are not shown for simplicity.}
\end{figure}

In Fig.~\ref{fig:Branching_Ratio}, we present the branching ratios of $H^\pm$ (left panels) and $H$ (right panels)
in the PDG (upper panels) and CDF cases (lower panels)
over the finally allowed parameter points.
The results of the PDG-island correspond to separate groups of the points for the light $\mch/\mhh$ in the upper panels.
In the PDG-island, $\ch\to \tau\nu$ and $H\to \ttau$ have almost 100\% branching ratios.
The muon modes, $\ch \to \mu\nu$ and $H \to \mmu$,
have about 0.3\% branching ratios,
which are omitted to avoid congestion.
In the PDG-island, the bosonic decay modes are extremely suppressed
such that $\br(\ch\to A W^*) \lsim 1.1 \times 10^{-5}$ and $\br(H \to A Z^*) \lsim 3.5 \times 10^{-5}$.
In the mainland regions of the PDG and CDF cases,
the bosonic decay modes of $\ch$ and $H$ are dominant over the leptonic modes.
The minimum of $\br(H^\pm \to W^\pm A)$ is about 60\% (70\%) in the PDG (CDF) case.
And $\br(H\to ZA)$ is above about 60\% in both the PDG and CDF cases. 

Based on the branching ratios, 
we study the multi-$\tau$ states through the electroweak processes.
First, $3\tau+\nu$ states are from
\begin{eqnarray}
 \label{eq:3tau_channels}
3\tau +\nu: && p p \to  H^\pm A \to [\tau^\pm \nu_{\tau}] [\tau^+ \tau^-] ,\\ \nn
&& p p \to   H^\pm H \to [\tau^\pm \nu_{\tau}] [\tau^+ \tau^-]  .
\end{eqnarray}
The $4\tau$ states consist of
\begin{eqnarray}
\label{eq:4tau_channels}
4\tau: && p p \to   H A \to [\tau^+ \tau^-][\tau^+ \tau^-],  \\[5pt] 
\label{eq:4tauV}
4\tau +V: && p p \to  H^\pm A \to [W^\pm A]A \to [W^\pm \tau^+ \tau^-][\tau^+\tau^-] ,\\  \nonumber
&& p p \to  HA \to [ZA]A \to [Z \tau^+ \tau^-][\tau^+\tau^-],\\[5pt] \nonumber
4\tau +VV': && p p \to  H^\pm H \to [W^\pm A][ZA] \to [W \tau^+ \tau^-][Z \tau^+\tau^-] , \\ 
\label{eq:4tauVV}
&& p p \to  H^+ H^- \to [W^+ A][W^- A] \to [W^+ \tau^+ \tau^-][W^- \tau^+ \tau^-],
\end{eqnarray}
where $V^{(\prime)}=Z,\wpm$.
The production of $HA$ ($\ch A$), mediated by $Z$ ($\wpm$),
is favored by the Higgs alignment because the vertex of $Z$-$H$-$A$ ($\wpm$-$\ch$-$A$) is proportional to $\sba$.

To calculate the production cross sections of the multi-$\tau$ states, 
we first implement the type-X 2HDM in \textsc{FeynRules}~\cite{Alloul:2013bka} 
to obtain the Universal FeynRules Output (UFO)~\cite{Degrande:2011ua}.
Interfering the UFO file with \textsc{Madgraph5-aMC@NLO}~\cite{Alwall:2011uj},
we compute the cross-sections of $pp \to H^\pm A/H^\pm H/HA/H^+H^-$ at 14 TeV LHC using \textsc{NNPDF31\_lo\_as\_0118} \cite{NNPDF:2017mvq} parton distribution function set. 
The two-body cross-sections are multiplied by relevant branching ratios of $A$, $H^\pm$ and $H$
from the \textsc{2HDMC}~\cite{Eriksson:2009ws}.\footnote{The 2HDM \textsc{UFO} file in the \textsc{MadGraph} misses some important decay modes of BSM scalar bosons
such as $\ch\to cs$ and $A \to gg$.}

\begin{figure}[t] \centering
\begin{center}
\includegraphics[width=\textwidth]{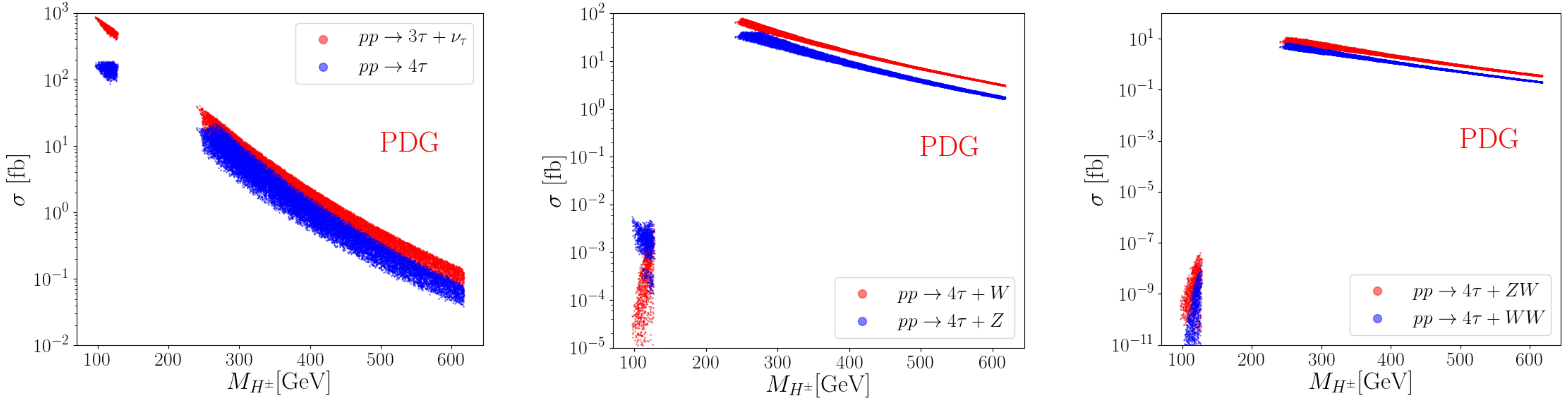}\\
\,\includegraphics[width=\textwidth]{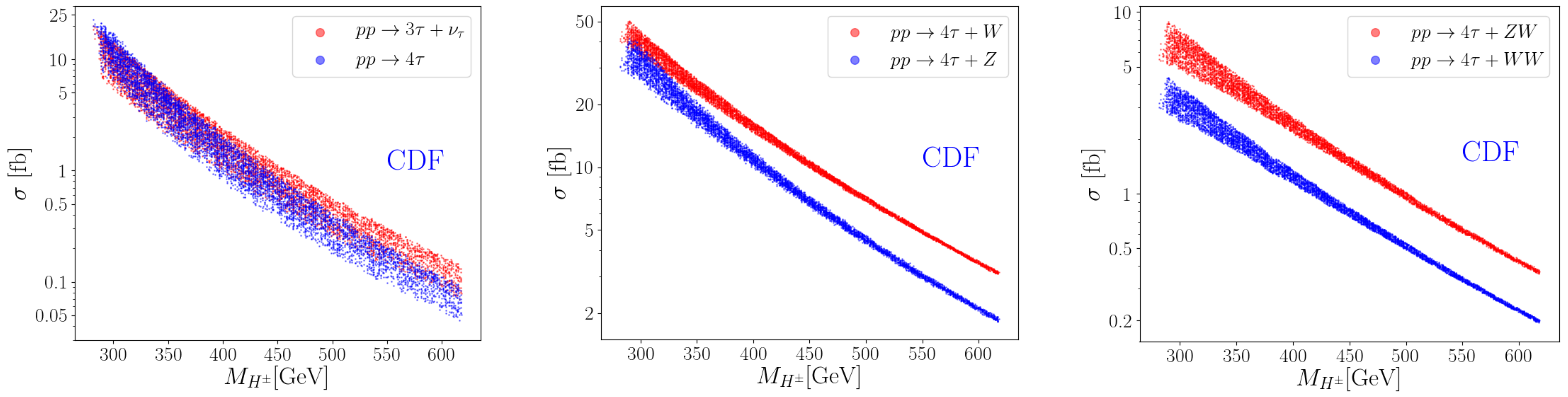}
\end{center}
\caption{\label{fig:cross_sections}
Production cross-sections of multi-$\tau$ states as a function of $\mch$:
$3\tau/4\tau$ (left panel), $4\tau + W(Z)$ (middle panel), and $4\tau + ZW(WW)$ (right panel). 
The PDG (CDF) results are in the upper (lower) panels. }
\end{figure}

Figure \ref{fig:cross_sections} presents the parton level cross-sections of $3\tau$ and $4\tau$ states in Eqs.~(\ref{eq:3tau_channels}) and  (\ref{eq:4tau_channels}).  
We compare the PDG results (upper panels) with the CDF results (lower panels).
The left panels show the cross-sections of the $3\tau$  and $4\tau$ states without a gauge boson.
In the middle (right) panels, we show the cross-sections of $4\tau + V$ ($4\tau + VV'$).
The PDG-island, which corresponds to $\mch \lsim 128\gev$ in the upper panels,
shows different behaviors:
the cross sections of $3\tau$ and $4\tau$ are substantially large, of the order of $1\pb$ and $100\fb$ respectively;
the cross sections of $4\tau+V$ and $4\tau+VV'$ are highly suppressed
like $\sg(pp\to 4\tau+VV')\lsim 10^{-7}\fb$.
It is attributed to the similar masses of BSM Higgs bosons as in \eq{eq:PDG-island},
which suppress the bosonic decays.
So, $3\tau$ and $4\tau$ states are the golden modes for the PDG-island.
On the other hand, 
the PDG-mainland yields a similar signal rates to the CDF.
The cross sections of $4\tau + V$ is several times larger than those of $3\tau/4\tau$ 
due to the dominant bosonic decays of $H^\pm$ and $H$.
The cross-sections of $4\tau + VV'$ are a few times smaller than those of $4\tau + V$.

\begin{figure}[t] \centering
\begin{center}
\includegraphics[width=0.9\textwidth]{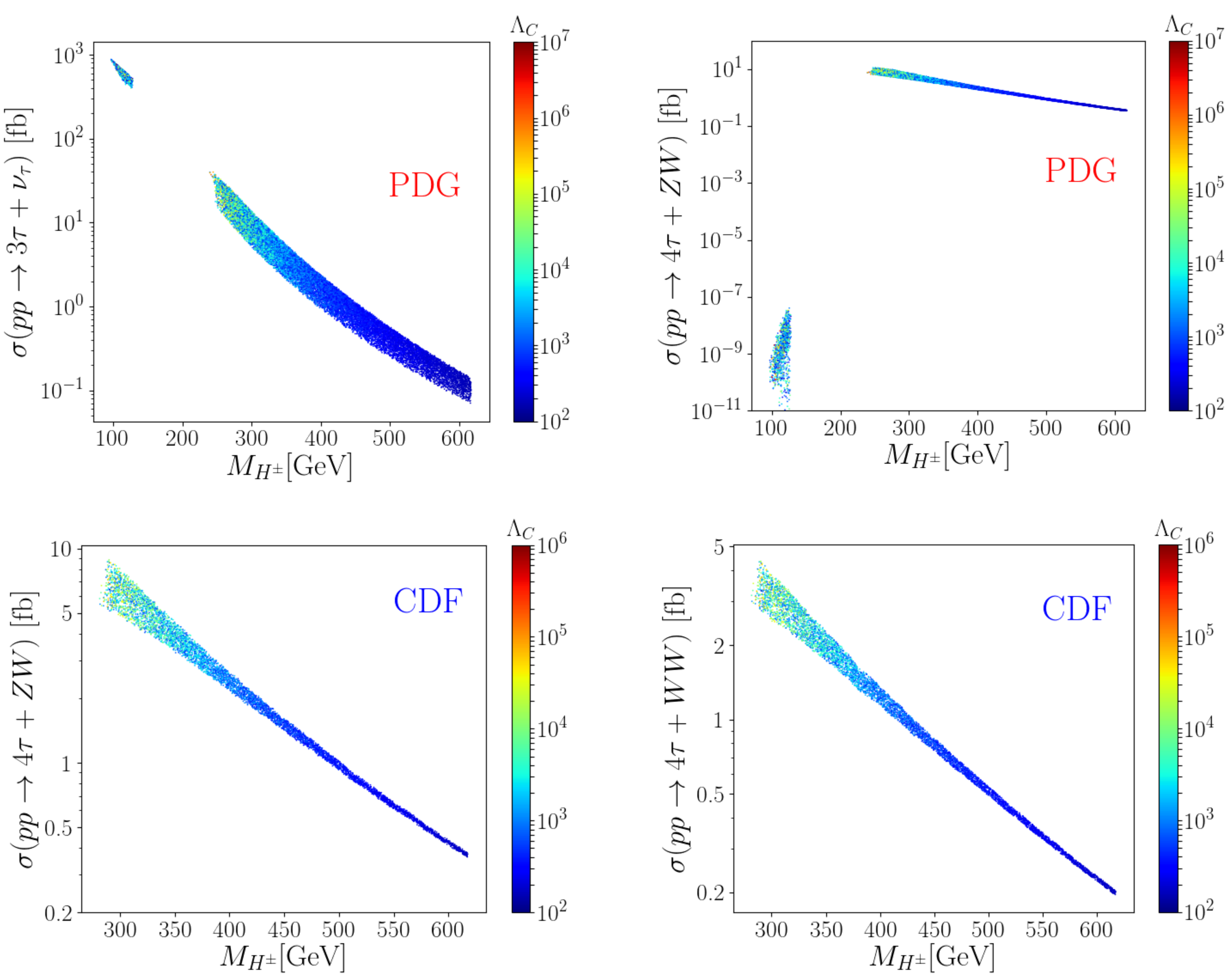}
\end{center}
\caption{\label{fig:xsections:cutoff}
The correlation between the cutoff scale $\lmc$ and the signal rates.
In the PDG case,
we present $\sg(pp\to3\tau+\nu)$ in the upper-left panel and $\sg(pp\to4\tau+ZW)$ in the upper-right panel,
as a function of $\mch$.
In the CDF case, we show $\sg(pp\to 4\tau+ZW)$ in the lower-left panel and $\sg(pp\to 4\tau+WW)$ 
in the lower-right panel.
The color codes indicate the cutoff scale $\lmc$.
}
\end{figure}

Now we discuss the correlation of the signal rates to the cutoff scales.
Among six processes in Eqs.~(\ref{eq:4tau_channels}), (\ref{eq:4tauV}),
and (\ref{eq:4tauVV}),
we concentrate on $3\tau+\nu$ and $4\tau+ZW$ for the PDG case
while $4\tau+ZW$ and $4\tau+WW$ for the CDF case.
The $3\tau+\nu$ state targets the PDG-island.
In Fig.~\ref{fig:xsections:cutoff},
we present the cross sections as a function of $\mch$ with the color code indicating $\lmc$.
The PDG results are in the upper panels, and the CDF results are in the lower panels.
The color codes clearly show that all four processes have maximal signal rates when the cutoff scale is large.
This correlation to $\lmc$ has a remarkable implication on the LHC phenomenology, such that the more valid the model is, the higher the discovery potential at the LHC is.

Based on the results in Figs.~\ref{fig:cross_sections} and \ref{fig:xsections:cutoff}, 
we propose $4\tau + VV'$ as the golden channel to probe the Higgs-phobic type-X.
First, the process, if observed at the HL-LHC, can exclude the PDG-island.
The second merit is that the higher cutoff scale guarantees the larger cross section. 
The most important merit of  $4\tau + VV'$ is almost background-free environment.
For the irreducible backgrounds, we calculate the parton level cross sections of 
$4\tau+Z\wpm$ and $4\tau+W^+ W^-$ in the SM by using the \textsc{MadGraph5-aMC@NLO}~\cite{Alwall:2011uj}.
We minimally impose the kinematic cuts on $\tau$ as 
$p_T^\tau>10\gev$, $|\eta_\tau|<2.5$, and $\Delta R(\tau,\tau)>0.4$.
The SM cross sections are $\sg(pp\to 4\tau+Z\wpm) \simeq 0.26\,{\rm ab}$
and $\sg(pp\to 4\tau+W^+ W^) \simeq 0.54\,{\rm ab}$,
which are negligible.
Reducible backgrounds are the production of four QCD jets plus $Z\wpm$ or $W^+ W^-$, 
where the QCD jets are misidentified as 
hadronically decaying tau lepton, $\tau_{\rm h}$.
Considering the mistagging rates of $P_{j \to \tau_{\rm h}}  =0.02$ in the one-prong decays
and $P_{j\to \tau_{\rm h}} =0.01$ in the three-prong decays,
it is hard for the QCD jets to mimic the $4\tau$ states. 
In addition, the large missing transverse energy cut additionally helps to tame the QCD jet backgrounds.
Other possible reducible backgrounds would be $t\bar{t}$+jets, $V$+jets, and  $VV'$+ jets.
We can significantly reduce $V(V')$+jets backgrounds by imposing the selection cuts like 
$n_\ell \geq 2$ and $n_{\tau_{\rm h}} \geq 4$, where $\ell = e, \mu$. 
In addition to that, the $b$-veto will kill the $\ttop+$jets background.

Nevertheless, there are concerns about the tau tagging.
Due to the low mass of $A$ and the decay chains involving $W^\pm$ and $Z$, 
the $\tau$-jets will be soft, which results in a low $\tau$-tagging efficiency. 
In that situation, an alternative would be to consider the mixed state like 
$2\ell+2\tau_{\rm h}$ decay mode of $4\tau$~\cite{Kanemura:2011kx}. 
Then the final state of $2\ell+2\tau_{\rm h}+Z\wpm (W^+ W^-)$ 
with leptonic decays of $Z$ and $\wpm$ results in five (four) $\ell$'s,
of which the backgrounds are negligible.
In the era of the new $W$ boson mass and the persistent $\damu$, 
it is worth studying the feasibility of $4\tau+VV'$ states at the high luminosity phase of LHC, which we leave for our future study.   

\section{Conclusions}
\label{sec:conclusions}

The recent measurement of the $W$ boson mass by the CDF collaboration
requests new physics beyond the SM: the Peskin-Takeuchi parameters 
significantly deviate from the SM expectation like
$S_{\rm CDF} = 0.15 \pm 0.08$ and $T_{\rm CDF}=0.27 \pm 0.06$ with $U=0$.
Another anomaly from the muon anomalous magnetic moment has been around for some time.
Type-X in the 2HDM is one of the most attractive solutions for the muon $g-2$ via a light pseudoscalar boson. 
Since the ordinary type-X suffers from $h\to AA$ and the lepton flavor universality data in the $\tau$ and $Z$ decays,
we have proposed the Higgs-phobic pseudoscalar in type-X.

Through random scanning of the model parameters,
we impose the theoretical and experimental constraints step by step:
step I is for the muon $g-2$ and theoretical stabilities;
step II is for the oblique parameters before and after the CDF $m_W$ measurement;
step III applies the Higgs precision data and the direct search bounds at high energy colliders;
step IV includes the global $\chi^2$ fit to $\damu$ and the LFU data.
The most important consequence is that the Higgs-phobic type-X can explain
not only $\mwcdf$ and $\damu$ anomalies
but also all the other constraints, including the LFU observables.

Our main results are summarized as follows:
\ben
 \item The muon $g-2$ anomaly requires light $\ma$ and large $\tb$.
  \item The PDG and CDF cases share some common features:
 	\bit
	\item The theoretical constraints and the electroweak oblique parameters put
	the upper bounds on $M_{H,\ch}\lsim 600\gev$.
	\item The LFU data plays the essential role in the curtailment of the parameter space,
  eliminating most of the region with $\ma\gsim 38\gev$ and $\tb \gsim 70$.
	\eit
\item There exist meaningful differences between the PDG and CDF cases:
 	\bit
	\item Only	in the PDG case,
	a small region where $\ma\simeq\mch\simeq\mhh\simeq100\gev$ 
	survives to the last step, called the PDG-island.
	\item The PDG-island accommodates both the right-sign and wrong-sign tau lepton Yukawa coupling,
	while outside the PDG-island only the wrong-sign $\tau$ Yukawa coupling is allowed.
		\item The lower bound on $\mch$ for $\ma\lsim 38\gev$ is different, $\mch\gsim 250\gev$ in the PDG case
	but $\mch\gsim 300\gev$ in the CDF case.
	\item The cutoff scale in the PDG case can go higher than in the CDF case,
	the former up to $10^7\gev$ and the latter to $10^5\gev$.
	\eit
 \item We propose the $4\tau$ states associated with $ZW$ or $WW$ as the golden discovery modes at the LHC for the CDF case,
 because of the background-free environment.

\een

\acknowledgments
The work of JK, SL, and JS is supported by 
the National Research Foundation of Korea, Grant No.~NRF-2022R1A2C1007583. 
The work of P.S. was supported by the appointment to the JRG Program at the APCTP through the Science and Technology Promotion Fund and Lottery Fund of the Korean Government. This was also supported by the Korean Local Governments - Gyeongsangbuk-do Province and Pohang City.

\appendix

\section{Contributions to $\damu$ in type-X}
\label{appendix:damu}
In the 2HDM,
there exist two kinds of contributions to $\damu$,
one-loop contributions 
and two-loop Barr-Zee contributions~\cite{Barr:1990vd,Ilisie:2015tra}.
The one-loop contributions mediated by $H$, $A$, and $\ch$ are~\cite{Chun:2016hzs}
\bea
\label{eq:amu:oneloop}
	\Delta a_\mu^{\rm 1-loop} &=&
	\frac{G_F \, m_{\mu}^2}{4 \pi^2 \sqrt{2}}  \, 
	\sum_\phi  \left( \xi_{\mu}^\phi \right)^2  \rho^\mu_\phi \, 
	f_\phi(\rho^\mu_\phi),
\eea
where $\phi =  \{H, A , H^\pm\}$ and
$\rho^i_j=  m_i^2/m_j^2$.
The loop function
$f_{\phi}$ is
\begin{align}
f_H(\rho) &= \int_0^1 d x \frac{x^2(2-x)}{1-x +\rho x^2},
\\ \nn
f_A(\rho) &= -\int_0^1 d x \frac{x^3}{1-x +\rho x^2},
\\ \nn
f_{\ch}(\rho) &= -\int_0^1 d x \frac{x(1-x)}{1- \rho(1-x)}.
\end{align}

At two-loop level,
dominant contributions are from the Barr-Zee type diagrams with heavy fermions 
in the loop, given by~\cite{Barr:1990vd}
\bea
\label{eq:BZ}
	\Delta a_\mu^{\rm BZ} 
	&=& \frac{G_F \, m_{\mu}^2}{4 \pi^2 \sqrt{2}} \, \frac{\alpha_{\rm em}}{\pi}
	\, \sum_{f,\phi^0}  N^c_f  \, Q_f^2  \,  \xi_{\mu}^{\phi^0}  \, \xi_{f}^{\phi^0} \,  \rho^{f}_{\phi^0}\,  
	g_{\phi^0}(\rho^{f}_{\phi^0})
	,
\eea
where $f=t,b,\tau$, $\phi^0 = H,A$,  $m_f$, $Q_f$ and $N^c_f$ are 
the mass, electric charge and color factor of the fermion $f$.
The loop functions are
\bea
\label{eq:2loop:ft}
	g_{H}(\rho) &=& \int_0^1 \! dx \, \frac{2x (1-x)-1}{x(1-x)-\rho} \ln \frac{x(1-x)}{\rho},
	\\[3pt] \nn
	g_A(\rho) &=& \int_0^1 \! dx \, \frac{1}{x(1-x)-\rho} \ln \frac{x(1-x)}{\rho}.
\eea
For light $\ma$ and $\tb\gsim 30$,
the largest contribution is from the Barr-Zee diagram with $\tau$ loop,
mediated by $A$.

\section{Lepton flavor universality observables in the 2HDM}
\label{appendix:LFU}

For the HFLAV global fit results in the $\tau$ decay,
the coupling ratios in the 2HDM are\footnote{Equation (29) in Ref.~\cite{Jueid:2021avn} has a typo. Correct one is $\mathcal{R}^\tau_{1,2,3}=1+\dt_{\rm loop}$.}
\bea
\frac{g_\tau }{ g_\mu} &=&  \lf  g_\tau \over g_\mu \ri_\pi = \lf  g_\tau \over g_\mu \ri_K=1+\dt_{\rm loop},
\\[3pt] \nn
\frac{g_\tau}{ g_e} &=& 1+ \dt_{\rm loop}+ \es^\tau_{\rm tree} ,
\\[3pt]
 \frac{g_\mu}{ g_e}  &=& 1+ \es^\tau_{\rm tree},
\eea
where $\dt_{\rm loop}$ and $\es^\tau_{\rm tree}$ are
\bea
\label{eq:dt:loop}
\dt_{\rm loop} &=& \frac{1}{16\pi^2} \frac{m_\tau^2\tb^2}{v^2}
\left[
1+\frac{1}{4} \left\{
k\lf \rho^A_{\ch} \ri + k \lf \rho^{H}_{\ch} \ri
\right\}
\right],
\\ \label{eq:es:tree}
\es^\tau_{\rm tree} &=& \dt_{\rm tree} 
\left[\frac{ \dt_{\rm tree}}{8}
-\frac{m_\mu}{m_\tau} \frac{g\lf \rho^\mu_\tau \ri}{f\lf \rho^\mu_\tau \ri}
\right].
\eea
The expression in \eq{eq:dt:loop}  is valid 
in the Higgs alignment limit, which is almost maintained in our model.
Here $\dt_{\rm tree} $ denotes the generic tree-level contribution mediated by the charged Higgs boson,
given by
\bea
\dt_{\rm tree} = \frac{m_\mu m_\tau \tb^2}{\mch^2}.
\eea
The loop functions in \eq{eq:dt:loop}  and \eq{eq:es:tree} are
\bea
k(x) &=& (1+x)\ln x/(1-x),
\\ \nn
g(x) &=& 1+9x -9 x^2-x^3 + 6 x (1+x) \ln x,
\\ \nn
f(x) &=& 1-8x+8x^3-x^4 -12 x^2 \ln x.
\eea

Among the Michel parameters in \eq{eq:Michel:def},
the 2HDM only affects $\eta_\mu$, $ \lf \xi \dt \ri_\mu$, and $\xi_\mu$ as
\bea
\label{eq:Michel:2HDM}
\eta_\mu &=& - \frac{2  \dt_{\rm tree}  (1+\dt_{\rm loop})}{4 + \dt_{\rm tree}^2},
\\[3pt] 
\label{eq:Michel:xidt}
\lf \xi \dt \ri_\mu &=&
\frac{3}{4}\times
\frac{4(1+\dt_{\rm loop})^2-\dt_{\rm tree}^2}{4(1+\dt_{\rm loop})^2+\dt_{\rm tree}^2},
\\[3pt] 
\label{eq:Michel:xi}
\xi_\mu &=& \frac{4(1+\dt_{\rm loop})^2-\dt_{\rm tree}^2}{4(1+\dt_{\rm loop})^2+\dt_{\rm tree}^2}.
\eea
And the new contributions to the leptonic $Z$ decays are written as
\bea
\frac{\Gm(Z\to l^+ l^- )}{\Gm(Z \to \ee)} -1  = 
\frac{2 g_L^\sm {\rm Re}\lf \dt g_{L}^l \ri +2 g_R^\sm {\rm Re}\lf \dt g_{R}^l \ri}
{\lf g_L^\sm\ri^2+\lf g_L^\sm\ri^2},\quad \lf l=\mu,\tau \ri
\eea
where $g_L^\sm=s_W^2-1/2$, $g_R^\sm=s_W^2$,
and the full expressions for $\dt g_{L/R}^{\mu,\tau}$ at one-loop level are referred to Ref.~\cite{Chun:2016hzs}.


\end{document}